\documentclass[conference]{IEEEtran}
\IEEEoverridecommandlockouts

\usepackage{cite}
\usepackage{amsmath,amssymb,amsfonts}
\usepackage{graphicx}
\graphicspath{{./}}
\usepackage{textcomp}
\usepackage{xcolor}
\usepackage[hyphens]{url}

\newcommand{\revA}[1]{#1}
\newcommand{\revB}[1]{#1}
\newcommand{\revC}[1]{#1}
\newcommand{\revD}[1]{#1}
\newcommand{\revE}[1]{#1}
\newcommand{\revF}[1]{#1}
\newcommand{\revMulti}[1]{#1}
\newcommand{\revShep}[1]{#1}

\usepackage{amsfonts,multirow, mathtools,bm}
\usepackage{enumitem}
\usepackage{stfloats}
\usepackage{algorithm}
\usepackage{algpseudocode}
\usepackage[caption=false,font=footnotesize]{subfig}
\usepackage{braket}
\usepackage{booktabs}
\newcommand{\XW}[1]{}
\def\BibTeX{{\rm B\kern-.05em{\sc i\kern-.025em b}\kern-.08em
    T\kern-.1667em\lower.7ex\hbox{E}\kern-.125emX}}
\begin{document}

\pdfpagewidth=8.5in
\pdfpageheight=11in
\title{Triage: An Adaptive Parallel Window Decoding Scheduler for Real-time Fault-Tolerant Quantum Computation
\thanks{$^*$Co-first authors}
\thanks{$^\dagger$Co-corresponding authors}
\thanks{This work has been partially supported by the National Key R\&D Program of China (Grant No.~2024YFB4504004), the National Natural Science Foundation of China (Grant No.~12447107), the Guangdong Provincial Quantum Science Strategic Initiative (Grant Nos.~GDZX2403008 and GDZX2503001), and the Guangdong Provincial Key Lab of Integrated Communication, Sensing and Computation for Ubiquitous
Internet of Things (Grant No.~2023B1212010007).}
}
\author{
\IEEEauthorblockN{1\textsuperscript{st} Jiahan Chen$^*$}
\IEEEauthorblockA{
\textit{The Hong Kong University of} \\
\textit{Science and Technology (Guangzhou)} \\
Guangzhou, China}
\and
\IEEEauthorblockN{2\textsuperscript{nd} Chenghong Zhu$^*$}
\IEEEauthorblockA{
\textit{The Hong Kong University of Science} \\
\textit{ and Technology (Guangzhou)} \\
Guangzhou, China}
\and
\IEEEauthorblockN{3\textsuperscript{rd} Ge Bai$^\dagger$}
\IEEEauthorblockA{
\textit{The Hong Kong University of} \\
\textit{Science and Technology (Guangzhou)} \\
Guangzhou, China \\
gebai@hkust-gz.edu.cn}
\and
\IEEEauthorblockN{4\textsuperscript{th} Xin Wang$^\dagger$}
\IEEEauthorblockA{
\textit{The Hong Kong University of Science} \\
\textit{\centerline{and Technology (Guangzhou)}}\\
Guangzhou, China \\
felixxinwang@hkust-gz.edu.cn}
}

\maketitle


\begin{abstract}

Fault-tolerant quantum computation (FTQC) critically depends on real-time classical decoding, which is rapidly emerging as a system bottleneck. As quantum systems scale, decoding latency and throughput limitations lead to exponential syndrome backlogs and logical operation stalls. \revA{While hardware accelerators and parallel windowing offer pathways to speed up decoding, dynamically deploying a finite pool of decoders across a vast quantum error correction architecture remains an unresolved resource allocation problem}.

\revMulti{To address this, we formulate FTQC decoding as a constrained dynamic scheduling problem by utilizing a spatio-temporal framework based on \emph{slices}. We propose Triage, a dual-mode architecture that mitigates operation stalls by adaptively combining a cost-efficient heuristic scheduler with a priority-aware emergency mode to rapidly resolve the causal cone of critical operations}. Our evaluation shows that Triage maintains low algorithm stalls and logical error rates even under scarce classical resource constraints. Across various benchmarks, Triage achieves an average logical error rate reduction of 52.6\% compared to standard temporal parallelism, enabling an efficient classical control plane for scalable FTQC architectures.

\end{abstract}


\section{Introduction}

\begin{figure}[!t]
    \centering
    \includegraphics[width=1\linewidth]{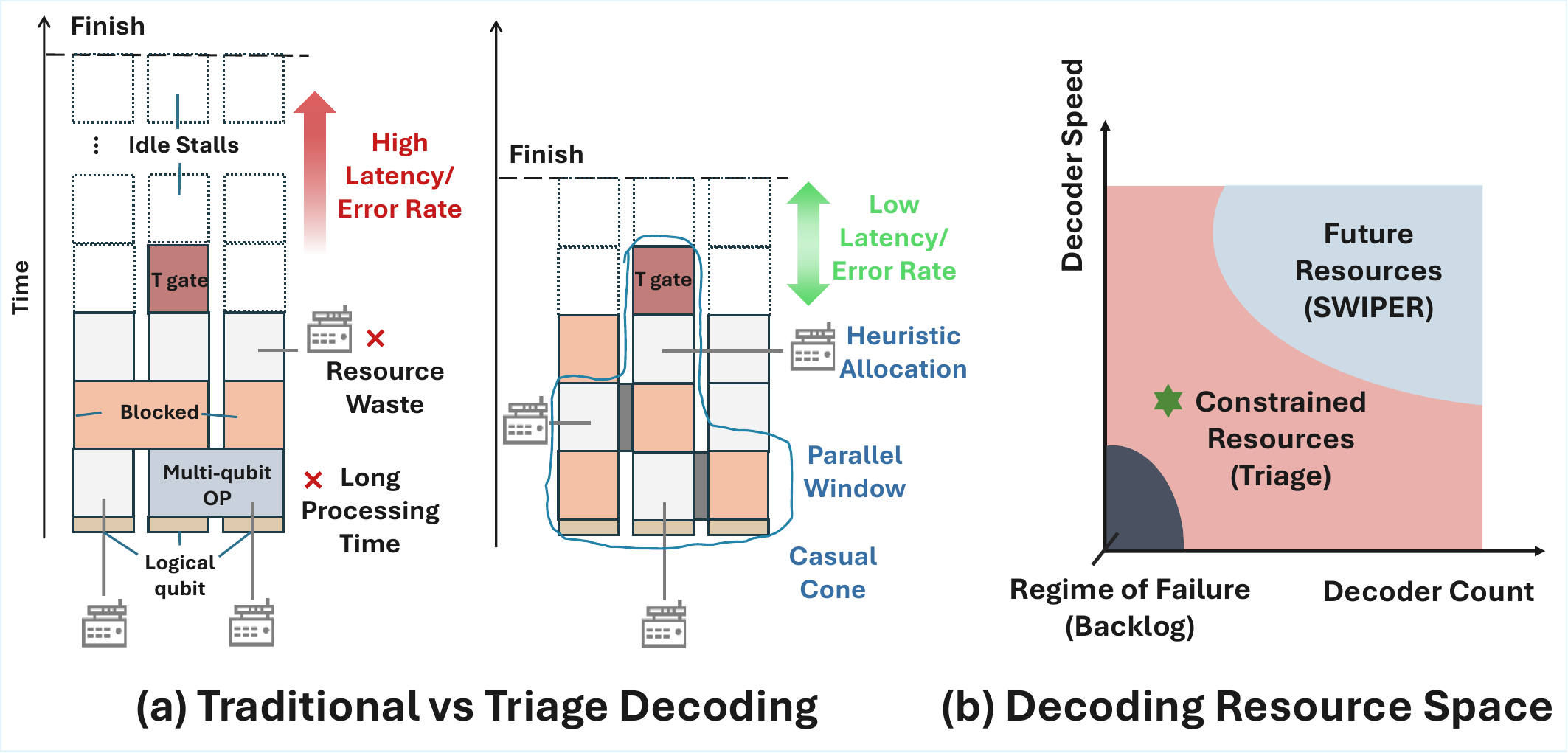}
    \caption{\revMulti{\textbf{Navigating the FTQC Decoding Bottleneck.} (a) Traditional decoding leads to large idle stalls, while Triage employs spatio-temporal windows and prioritizes the causal cone to effectively reduce the latency. (b) Triage achieves better performance in the near-term, resource-constrained landscape.}}
    \label{fig:teaser}
\end{figure}

Quantum computers hold the potential to efficiently solve certain problems that are intractable for the best known classical algorithms~\cite{shor1999polynomial, childs2018toward, harrow2009quantum}. However, current quantum hardware is highly error-prone~\cite{preskill2018quantum, clerk2010introduction}, requiring quantum error correction (QEC) to enable fault-tolerant quantum computation (FTQC)~\cite{campbell2017roads}. Encouragingly, recent experimental progress across various platforms and QEC codes~\cite{fowler2012surface, landahl2011fault, bravyi2024high} is rapidly validating this approach~\cite{google2025quantum, bluvstein2024logical, wang2025demonstration, caune2024demonstrating, eickbusch2024demonstrating}. As these advances move FTQC from theory toward viability, the focus shifts from theoretical feasibility to the architectural challenges of implementation.

Towards a large scale FTQC, a significant architectural bottleneck arises from the classic control layer. At the heart of this layer is the \emph{decoder}, whose function is to continuously process a massive stream of classical syndrome data from the quantum processor, and infer the most likely errors. Crucially, decoding must operate in \emph{real-time}. This means the overall throughput of decoding must, on average, exceed the rate of syndrome generation. Otherwise, the system will accumulate an exponential backlog~\cite{terhal2015quantum} of unprocessed syndromes, which will eventually overwhelm the computational resources.

To address this challenge, one line of research focuses on optimizing the \emph{latency} of a single decoding operation. On the software front, significant effort has gone into developing high-accuracy decoding algorithms with low computational complexity~\cite{higgott2025sparse, delfosse2022toward, chen2025improved, muller2025improved}. In parallel, hardware accelerators using FPGAs have demonstrated single-decode latencies below the demands of superconducting qubits for certain code distances~\cite{das2022lilliput, vittal2023astrea, alavisamani2024promatch, wu2025micro}. However, these hardware demonstrations have been largely confined to memory experiments rather than integrated into logical computations. Meanwhile, the protocol that processes the syndrome stream in a \emph{serial} fashion has been shown to be unscalable~\cite{skoric2023parallel}. Therefore, beyond optimizations for latency, designing superior decoding protocols that enhance overall \emph{throughput} is essential.

A second direction aims to improve decoding throughput via \emph{parallelism}. Temporal parallelism, for instance, partitions the syndrome stream into time blocks, allowing concurrent processing of the non-adjacent blocks~\cite{skoric2023parallel, tan2023scalable, viszlai2025swiper}. For multi-qubit logical operations such as lattice surgery, spatial parallelism can also be employed by partitioning the syndromes from the involved regions~\cite{lin2025spatially}. In principle, a spatio-temporal parallel approach would allow the system to scale its total throughput simply by adding more decoder units. Nevertheless, \revA{while temporal or spatial windowing techniques are known, a systematic scheduling framework integrating resource-aware temporal and spatial parallelism has not yet been demonstrated}.

How to deploy a finite pool of decoders onto a FTQC application? First, there is an \emph{asymmetry} between classical resources and logical qubits. Depending on hardware limitations or error targets, an application may require a few high-speed decoders for small codes, or a large pool of decoders collaborating in parallel for large codes. \revD{In a realistic large-scale architecture, this necessitates an $M$-for-$N$ shared resource model where $M < N$ decoders must be dynamically allocated.} At any moment, determining \emph{which} decoder to assign to \emph{which} logical patch is a resource allocation problem~\cite{maurya2024managing}.

This scheduling problem is further complicated by the operational logic of FTQC. The \emph{Pauli frame}~\cite{riesebos2017pauli} stores the decoder's inference of accumulated errors. When the computation encounters a non-Clifford gate, the decoder must update all relevant Pauli frames, a process we term \emph{synchronization}. A synchronization failure forces the logical operation to stall. During this idle period, qubits undergo additional error correction rounds, directly increasing the logical error rate (LER). To maximize fidelity via idle-reduction, decoding tasks relevant to the critical non-Clifford operation must be given higher priority. Combining this urgency with the spatio-temporal dependency constraints from parallel decoding, the problem is transformed into a dynamic constrained scheduling problem. \revMulti{As illustrated in Figure~\ref{fig:teaser}, traditional decoding approaches fail to navigate these dependencies efficiently, leading to severe resource waste and long idle stalls. Furthermore, as the hardware design space spans diverse decoder speeds and counts, a robust scheduling strategy becomes critical, especially in near-term, resource-constrained environments where naive policies quickly fall into an unrecoverable regime of failure.}

In this paper, we systematically address this challenge. First, \revA{we utilize a parallel spatio-temporal decoding framework using \textit{slice} (a $d \times d$ patch over $d$ rounds) as the basic scheduling unit}. By modeling the lifecycle of each slice and identifying the \emph{causal cone} of critical operations, we formulate the FTQC decoder scheduling problem. Second, \revA{to optimize performance under constrained resources}, we propose Triage, a dual-mode scheduling architecture which combines a fast heuristic-based steady mode with a robust look-ahead emergency mode. Triage significantly reduces the logical operation stalls, leading to an average LER reduction of 52.6\% compared to the standard temporal-parallel scheduling strategy.

In summary, we make the following contributions:
\begin{itemize}
    \item We introduce an abstraction of the decoder scheduling problem \revA{based on} a constraint graph of \textit{slices}. This framework is hardware-agnostic and applicable to diverse quantum platforms \revC{utilizing surface codes}.
    \item We propose the Triage scheduler, a dual-mode system that minimizes logical operation stalls by dynamically invoking an emergency mode to rapidly resolve the causal cone of prerequisite decodes. 
    \item We demonstrate that by effectively scheduling parallel windows, it is possible to overcome the latency limitations of individual decoders, enabling FTQC even in the challenging slow-decoder regime ($\tau_{decode} > \tau_{syndrome}$).
    \item \revE{We quantify the impact of real-time scheduling on system-level fidelity}, presenting a simulation framework that captures the interaction between syndrome generation and decoding.
\end{itemize}

\section{Background}

\subsection{Preliminary}

\subsubsection{Quantum Computing and Quantum Error Correction}

The fundamental unit of quantum computing is the qubit. A qubit inhabits a 2-D Hilbert space with computational basis states $\ket{0}$ and $\ket{1}$, and an arbitrary pure state can be written as $\ket{\psi} = \alpha\ket{0} + \beta\ket{1}$, where $\alpha$ and $\beta$ are complex amplitudes satisfying $|\alpha|^2 +|\beta|^2 =1$. Realistic qubits are noisy and error-prone. For example, a bit-flip error maps $\ket{\psi}$ to $\alpha\ket{1} + \beta\ket{0}$, and a phase-flip error maps $\alpha\ket{0} + \beta\ket{1}$ to $\alpha\ket{0} - \beta\ket{1}$. 
Quantum error correction (QEC) codes are necessary to preserve quantum information against such errors.

Stabilizer codes form a broad family of QEC codes that includes many of the most widely used constructions. A stabilizer code is specified by a set of commuting stabilizer generators $S_1, \ldots, S_m$, each of which is a Pauli operator acting as a tensor product of Pauli strings on the physical qubits. During syndrome extraction, dedicated ancilla qubits interact with the data qubits to measure the parity of each stabilizer without collapsing the encoded state. 

\subsubsection{Surface Code and Lattice Surgery}

\begin{figure}[ht]
    \centering
    \includegraphics[width=0.9\linewidth]{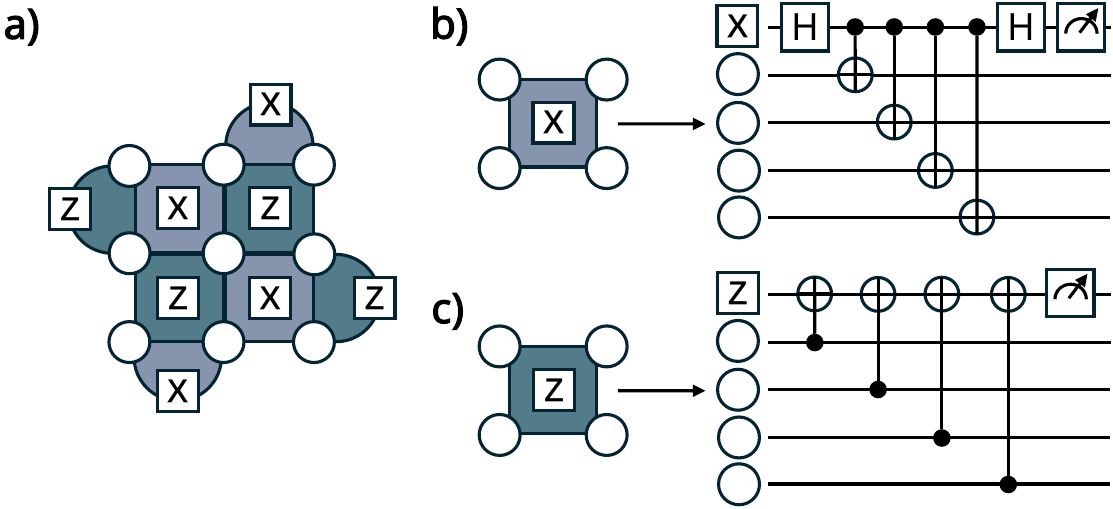}
    \caption{Example rotated surface code of distance $d = 3$. a) The code is defined by a set of X- and Z-type stabilizer checks used for syndrome extraction. b) and c) Syndrome extraction circuits for the X and Z stabilizers, respectively.}
    \label{fig:SC_syndrome_extraction_circuit}
\end{figure}

The surface code~\cite{fowler2012surface} has emerged as a leading candidate for building practical fault-tolerant quantum computers due to its high error threshold and hardware compatibility. Fig.~\ref{fig:SC_syndrome_extraction_circuit}(a) shows a rotated surface code of distance d = 3. The corresponding syndrome extraction circuit is shown in Fig.~\ref{fig:SC_syndrome_extraction_circuit}(b)(c): circles represent data qubits, and each data qubit couples to adjacent X- and Z-type ancilla qubits. The syndrome extraction is repeated over multiple rounds to collect measurement outcomes.

Lattice surgery~\cite{horsman2012surface} is a leading approach for implementing logical operations in surface code architectures. It works by measuring joint stabilizers along the boundaries of adjacent code patches, temporarily merging and then splitting patches~\cite{horsman2012surface, fowler2018low, de2020zx} to enact gate primitives. In contrast, non-Clifford operations such as the T gate, are typically supplied via magic state distillation or cultivation~\cite{bravyi2005universal, litinski2019magic, bravyi2012magic, jones2013multilevel, gidney2024magic}. In this distillation process, multiple noisy magic states are converted into fewer, higher fidelity states that are suitable for fault-tolerant state injection.

\begin{figure}[!ht]
    \centering
    \includegraphics[width=0.85\linewidth]{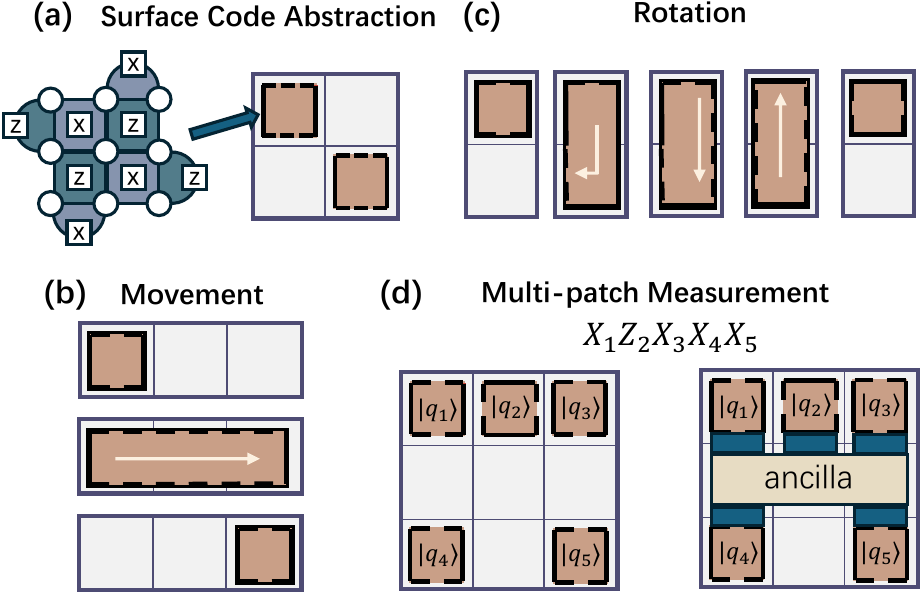}
    \caption{(a) Abstract view of the surface code as a patch. (b–d) Summary of the logical operations that can be performed.}
    \label{fig:LS_ops}
\end{figure}

Following the best practice~\cite{litinski2019game}, we represent each encoded surface-code qubit as a tile, as shown in Fig.~\ref{fig:LS_ops}(a). Building on this abstraction, Fig.~\ref{fig:LS_ops}(b–d) illustrates the key lattice surgery operations on tiles: patch movement, patch rotation, and multi-patch parity measurement. Execution of logical circuits typically follows the Pauli-Based Computation (PBC) paradigm, which systematically translates the universal Clifford+T circuits into a sequence of Pauli rotations. These rotations are then realized via the requisite lattice-surgery operations. We refer to~\cite{litinski2019game} for a detailed introduction.

\subsubsection{The Pauli Frame and T-Gate Synchronization}

The classical processing requirements for FTQC are fundamentally dictated by the Pauli frame~\cite{divincenzo2007effective, riesebos2017pauli} and its interaction with non-Clifford gates. The Pauli frame is a classical data structure that efficiently tracks the accumulation of Pauli errors on data qubits. This is a direct consequence of the definition of the Clifford group $\mathcal{C}_n$. If an operation $C \in \mathcal{C}_n$ is a Clifford gate, then for any Pauli operator $P \in \mathcal{P}_n$, the transformation results in another Pauli operator $P'=C P C^\dagger \in \mathcal{P}_n$. This property allows for efficient classical tracking: if an accumulated error $E \in \mathcal{P}_n$ exists on the state $|\psi\rangle$, applying a Clifford circuit $C$ transforms the state to $C(E|\psi\rangle) = (C E C^\dagger)(C|\psi\rangle) = E'(C|\psi\rangle)$. The new error $E'$ is also a Pauli operator and can be easily computed classically, allowing the frame to be updated without physical correction.

\begin{figure}[!ht]
    \centering
    \includegraphics[width=0.85\linewidth]{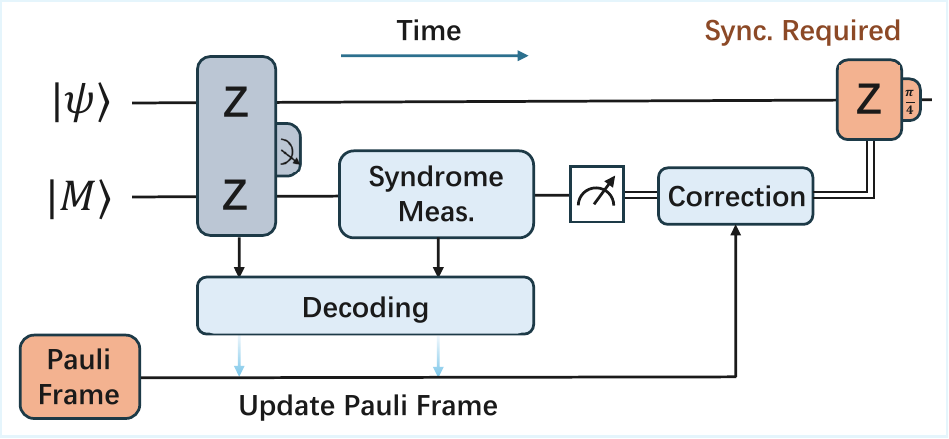}
    \caption{T-gate implementation via gate teleportation. The classically-controlled S-gate correction forces a decoder synchronization by physically correcting the Pauli frame.}
    \label{fig:T_gate_pauli_frame}
\end{figure}

This convenience ends, however, with the introduction of non-Clifford gates, such as the T-gate, which are essential for universal quantum computation~\cite{bravyi2005universal}. The T-gate breaks the classical tracking mechanism, as the transformed error can no longer be represented in the Pauli frame; for instance, $T X T^\dagger \notin \mathcal{P}_n$. As illustrated in Figure~\ref{fig:T_gate_pauli_frame}, a T-gate is typically implemented via preparing a high-fidelity magic state~\cite{litinski2019magic, itogawa2025efficient, gidney2024magic} and realizing a gate teleportation, which concludes with a measurement and a classically-controlled Pauli correction (an S-gate). Crucially, this final correction cannot be commuted through the T-gate and absorbed into the Pauli frame. Before this correction can be applied, the accumulated error on the logical qubit, $E_{acc}$, must be physically corrected by applying $E_{acc}^\dagger$. Only after the state is restored from $E_{acc}|\psi\rangle$ to $|\psi\rangle$ can the teleportation proceed correctly.

The central insight is the dichotomy in decoding requirements for FTQC. While the Pauli frame permits a relaxed, asynchronous approach to error correction, the presence of non-Clifford gates creates absolute synchronization points. They transform decoding into a priority scheduling problem, where synchronization failures lead to computational stalls and increased logical error rates.

\subsubsection{Window Decoding}

To manage the continuous stream of syndrome data in FTQC, decoders operate on discrete chunks of information known as windows. The traditional approach is \emph{serial sliding window} decoding~\cite{dennis2002topological}, where the temporal syndrome data is partitioned into fixed-size windows that are processed sequentially. However, this approach faces a scalability bottleneck. Let the time to generate the data for one window be $\tau_{gen}$ and the time for a single decoder to process it be $\tau_{dec}$. To prevent an exponential backlog of unprocessed syndromes, the system must satisfy the condition $\tau_{dec} < \tau_{gen}$~\cite{terhal2015quantum}. Assuming a decoder whose latency scales linearly with the number of qubits, $N$, i.e., $\tau_{dec} \propto N$, this constraint can be rewritten as:
\begin{equation}
    N < \frac{\tau_{round}}{k}
\end{equation}
where $\tau_{round}$ is the duration of a single syndrome measurement cycle and $k$ is a constant. This inequality reveals that for any decoder hardware, there exists an upper bound on the code distance that can be supported in real-time, rendering the approach unscalable for large QEC codes.

\begin{figure}[!ht]
    \centering
    \includegraphics[width=0.95\linewidth]{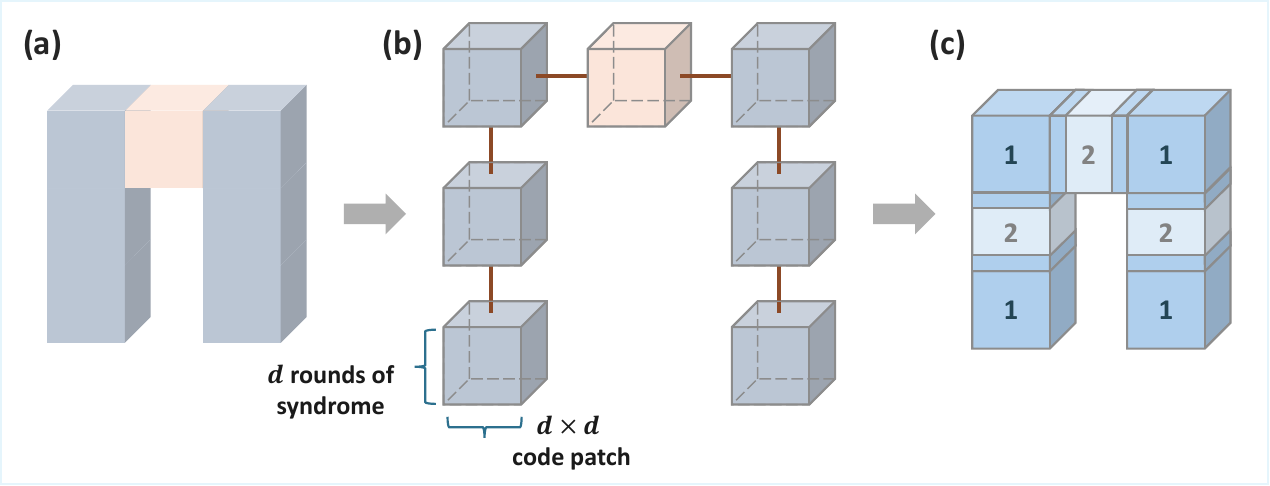}
    \caption{Spatio-temporal partitioning of a lattice surgery operation. (a) The monolithic operation volume. (b) It is decomposed into a graph of causally-constrained slices, where red edges represent mutual exclusion constraints. (c) The graph is 2-colored, which partitions all slices into two independent sets, and each set can be decoded in parallel.}
    \label{fig:window_decoding}
\end{figure}

The introduction of \emph{parallel window decoding} offered a solution. The key insight is that temporally disjoint windows are causally independent and can thus be decoded concurrently, as illustrated in Figure~\ref{fig:window_decoding}. \revC{In the time dimension, this allows for a checkerboard pattern of decoding, where all \emph{even} windows can be processed in parallel, followed by all \emph{odd} windows~\cite{tan2023scalable, skoric2023parallel}.} This concept extends naturally to the spatial dimension, where operations on different logical qubits can also be partitioned and processed in parallel~\cite{lin2025spatially}. Further refinements, such as speculative decoding~\cite{viszlai2025swiper}, aim to minimize the overhead at window boundaries. The parallel window approaches require that the decoding volume for a given window is expanded to include a buffer region containing syndrome data from its neighbors, with the window buffer size determining the extent of this look-ahead information. Then, the earlier decoded window creates artificial syndromes on its boundary with its neighbors. The insight of parallel window decoding is: with a sufficient number of parallel decoders, the system's overall throughput can be maintained even if individual decoders are slow ($\tau_{dec} \ge \tau_{gen}$).

\subsection{Motivation} 

\subsubsection{The Decoder Resource Dilemma}

Consider a surface code computation with a 2-D array of logical qubits, as shown in Figure~\ref{fig:LS_ops}. A real-time decoding system must continuously process the syndrome streams from all active qubit patches. How should we allocate resources in this classical system?

On one extreme, a \emph{one-to-one mapping} dedicates a physical decoder to each logical qubit. While maximizing parallelism, this approach is architecturally infeasible for large systems due to its prohibitive cost and resource underutilization. On the other extreme, a \emph{one-for-all approach} uses a constant number of decoders to service the entire machine. This is also unscalable, as the latency requirement for each decoder would scale inversely with the number of logical qubits ($O(1/N_{lq})$), an impossible demand for any non-trivial algorithm.

The only viable path is a shared-resource model: an \emph{M-for-N scheduler} that manages a pool of $M$ physical decoders to service the tasks from $N$ logical qubits, transforming the issue into a scheduling problem. The scheduler should make online decisions to prioritize tasks and maximize throughput, especially in two critical scenarios: the resource-constrained regime where fast decoders are scarce ($M \le N, \tau_{dec} < \tau_{gen}$), and the computationally-constrained regime where decoders may be individually slow ($M > N, \tau_{dec} \ge \tau_{gen}$).

Recent research has begun exploring this scheduling problem~\cite{maurya2024managing}. Existing models, based on a logical qubit level abstraction, offer a valuable first step but fail to capture the fine-grained complexity inherent in large-scale quantum algorithms. By shifting the scheduling abstraction down to the level of individual spatio-temporal windows, we can unlock two powerful dimensions of optimization. 

First, a fine-grained scheduler can effectively manage the strong \emph{spatial correlations} created by operations like lattice surgery. \revF{Because lattice surgery temporarily merges adjacent logical patches and measures joint stabilizers along their boundaries, errors become spatially correlated across the merged region. Instead of forcing a single decoder to serialize the decoding of this massive, combined volume, a slice-aware scheduler can partition it into smaller, spatially parallelizable windowed tasks}, simultaneously ensuring correctness and efficiency. Second, this approach can also fully leverage \emph{spatio-temporal parallelism} during non-Clifford operations. A slice-aware scheduler can dispatch multiple decoders to collaboratively resolve emergent windows, reducing synchronization latency. Furthermore, this decomposition is inherently more efficient, as it avoids the super-linear complexity penalty associated with decoding large, monolithic data blocks~\cite{higgott2025sparse}.

\subsubsection{The Scalability Crisis of Parallel Decoding}
\label{sec:The Scalability Crisis of Parallel Decoding}

Parallel window decoding~\cite{tan2023scalable, skoric2023parallel, viszlai2025swiper} have addressed the \emph{computational scalability} problem of sliding window decoding. However, this arises an architectural question: what is the practical upper bound on the number of required decoders $M$?

\begin{figure}[!ht]
    \centering
    \includegraphics[width=0.9\linewidth]{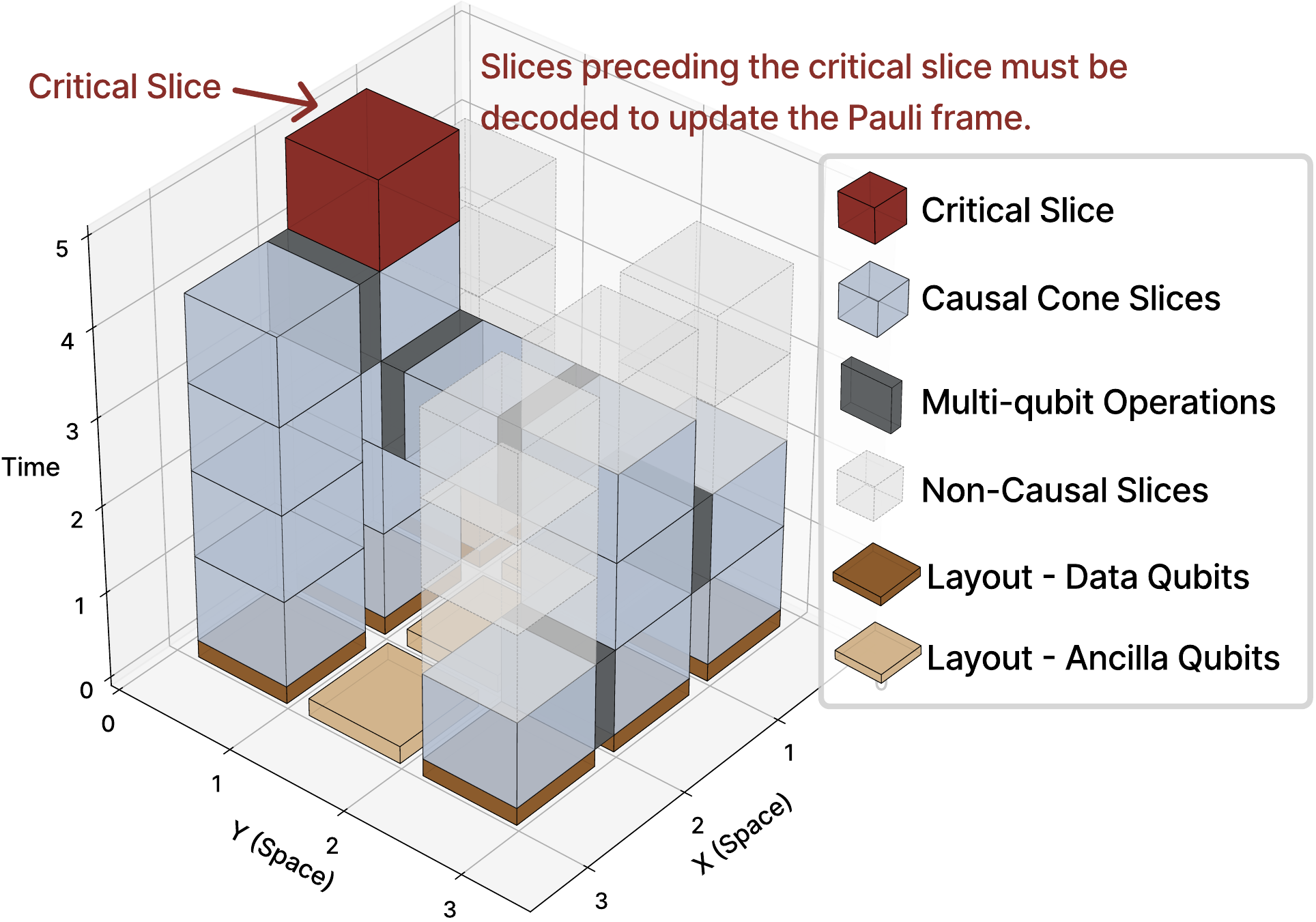}
    \caption{An illustration of the causal cone corresponds to one critical slice, which is the Clifford correction after the T-gate teleportation in Figure~\ref{fig:T_gate_pauli_frame}.}
    \label{fig:causal_cone}
\end{figure}

\revB{We define the complete set of dependencies for a critical operation as its \emph{causal cone}: the transitive closure of all un-decoded historical slices belonging to any logical qubit that has become correlated with the target through a chain of multi-qubit operations}, as illustrated in Figure~\ref{fig:causal_cone}. In the worst case, if the algorithm is highly entangled and the intervals between non-Clifford gates are long, this causal cone can grow to a large spatio-temporal volume. A brute-force parallel approach to resolve such a backlog at the last minute would require a number of decoders proportional to the backlog's size.

This reveals the \emph{resource scalability} as another potential crisis. The demand for decoders is highly non-uniform, with massive spikes preceding critical gates, making a static, worst-case provisioning of decoders architecturally infeasible. This insight is our central motivation: a dynamic online scheduler is essential to manage a finite decoder pool, handling the average workload efficiently while mobilizing maximum parallelism only when necessary to meet critical deadlines.

\section{Triage Scheduler}

We first define the scheduler system model and then present our Triage scheduling algorithm.

\subsection{System Model and Problem Formulation}
\label{sec:model}

\begin{figure*}[!ht]
    \centering
    \includegraphics[width=0.75\textwidth]{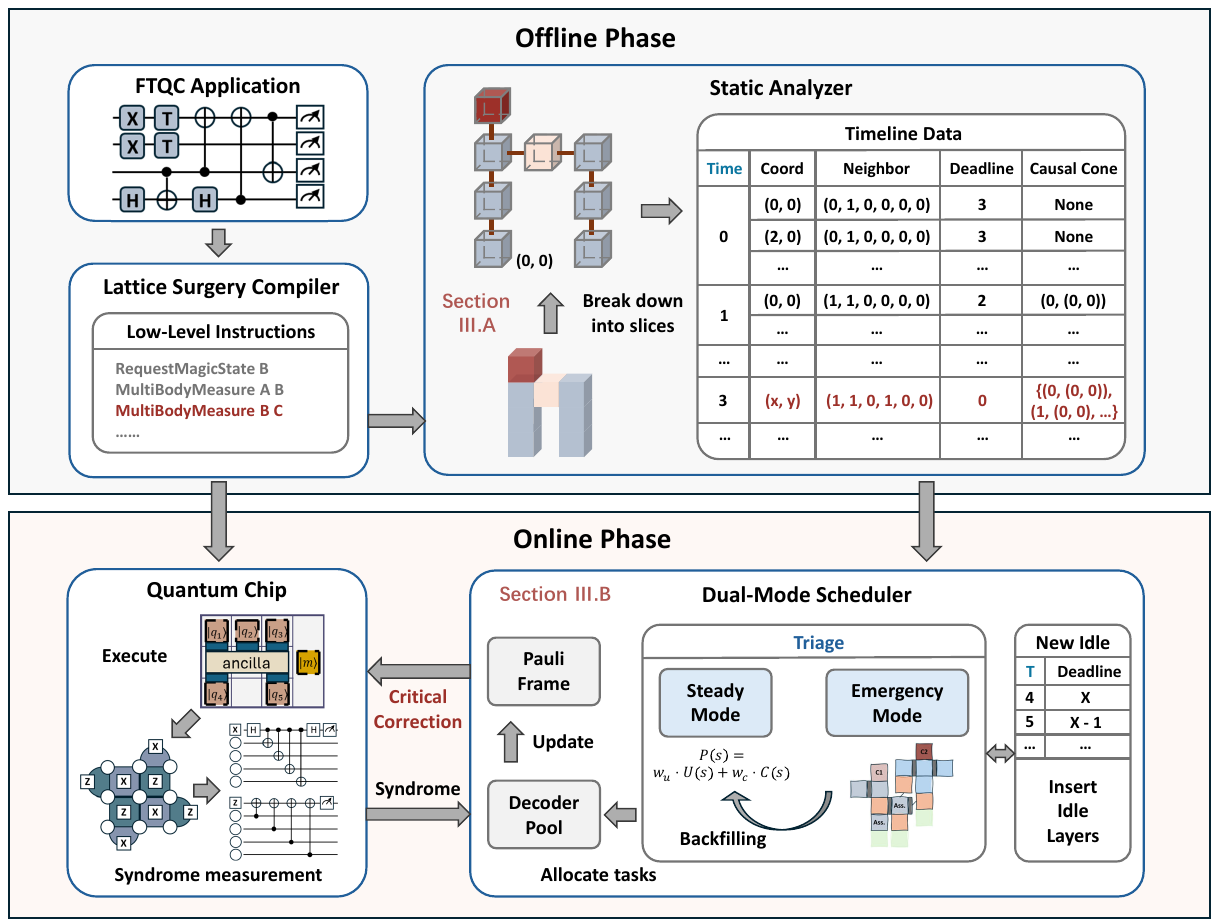}
    \caption{\revMulti{Architectural Overview of the Triage Scheduling Framework.} The offline phase consists of a compiler and a static analyzer that generate an annotated Timeline from LLIs. During the online phase, the scheduler uses this Timeline to dispatch a stream of syndrome data from the hardware to a finite pool of M physical decoders.}
    \label{fig:scheduler_architecture}
\end{figure*}

The decoder scheduler serves as middleware within the FTQC classical control stack. As illustrated in Figure~\ref{fig:scheduler_architecture}, its architecture is divided into offline and online phases.

\subsubsection{\revMulti{Offline Phase: Compile-Time Analysis}}

Due to the superlinear complexity of the decoder, processing a large block of syndrome data is less efficient than decoding its components individually. \revA{Building upon the concepts of parallel window decoding, we leverage fine-grained spatio-temporal partitioning to maximize the degree of parallelism. We therefore adopt the \emph{slice} as the atomic scheduling unit of our framework.}

A slice $S(t, p)$, represents the syndrome data generated from a single square logical patch at position $p$ during a $d$-rounds syndrome measurement cycle $t$. The scheduler models the computation as an undirected graph $G=(V, E)$, where the set of vertices $V$ consists of all slices. An edge $(u, v) \in E$ represents a mutual exclusion constraint signifying that slices $u$ and $v$ cannot be decoded concurrently. Each slice is characterized by a set of attributes that guide the scheduler's decisions:

\paragraph{Neighbors}
Edges in the constraint graph are defined by the neighbors of each slice. A slice $S(t, p)$ can have up to six neighbors: two for temporal predecessor ($t-1$) and successor ($t+1$), and four spatial neighbors at time $t$. A data qubit slice always has at least one temporal dependency due to the syndrome stream. Spatial dependencies are introduced by multi-qubit lattice surgery.

\begin{figure}[!ht]
    \centering
    \includegraphics[width=0.9\linewidth]{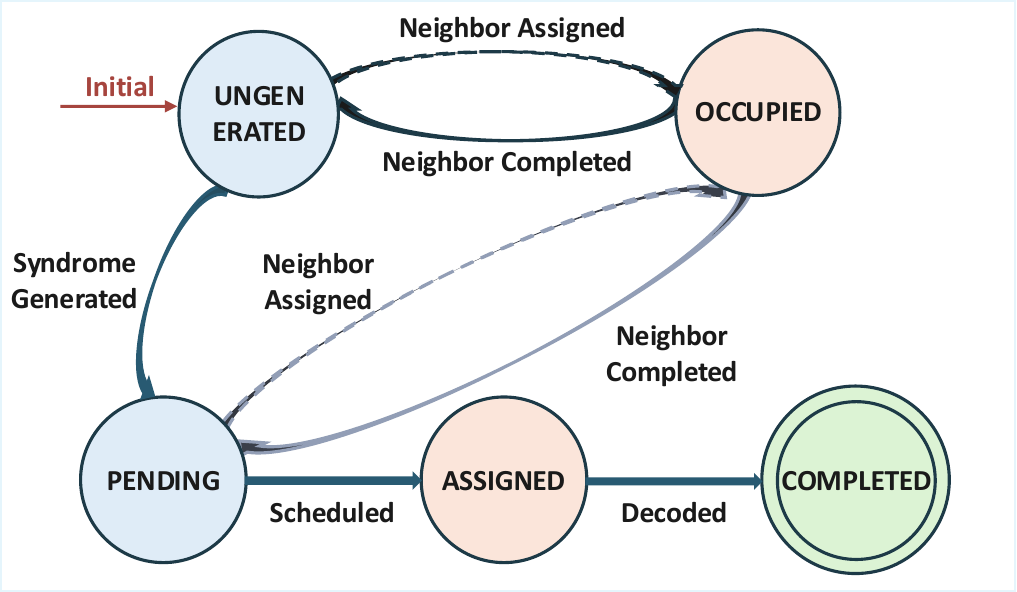}
    \caption{State transition diagram for a slice's lifecycle.}
    \label{fig:slice_lifecycle}
\end{figure}

\paragraph{Decoding Status}
Each slice maintains a state transitioning through the automata depicted in Figure~\ref{fig:slice_lifecycle}. A slice is initially \texttt{UNGENERATED}. Once its syndrome is produced by the quantum hardware, it becomes \texttt{PENDING}. A \texttt{PENDING} slice is eligible for decoding, but may be blocked by a neighbor that is currently being decoded; in this case, it is marked as \texttt{OCCUPIED}. When a slice is ready and selected by the scheduler, it moves to \texttt{ASSIGNED} for the duration of its decoding. Upon successful processing, it enters \texttt{COMPLETED}. The Timeline dynamically maintains \texttt{PENDING} and \texttt{COMPLETED} slices during the online phase to facilitate the scheduler's decision.

\paragraph{Causal Cone}
Slices associated with critical operations (i.e., Clifford corrections after a T-gate teleportation) have the causal cone as an additional attribute, which defines the complete set of historical slices that must be decoded to update the Pauli frame, as described in Section~\ref{sec:The Scalability Crisis of Parallel Decoding}. Our framework employs a lazy computation strategy. When the scheduler requires a causal cone, \revB{it is calculated on-demand via a backward BFS starting from the critical slice's immediate spatial and temporal predecessors (i.e., the slices directly participating in the non-Clifford gate). The traversal only expands to (i) same-layer spatial neighbors and (ii) the one-step temporal predecessor at $t-1$. A node in \texttt{COMPLETED} state will be pruned, while the BFS continues on the remaining frontier until the queue is exhausted.} The results are then stored in a bounded LRU cache. All subsequent queries related to the same critical operation are served instantly from the cache.

Based on the slice abstraction, the compiler first lowers a high-level program to a Low-Level Instruction (LLI) stream over a 2-D logical-qubit layout. \revB{A single pass then builds the \emph{Timeline} structure, where each unit stores: (i) an integer layer index \(t\) (unit: one syndrome-measurement cycle), (ii) spatial coordinate \((r,c)\), (iii) operation label, (iv) a 6-bit immediate-neighbor mask \((t\!-\!1,t\!+\!1,\uparrow,\downarrow,\leftarrow,\rightarrow)\), (v) a deadline measured in \emph{layers} to the nearest upcoming critical synchronization point (\(\infty\) if none), and (vi) a possible \emph{causal cone} reference.  
During online simulation, this discrete timeline is mapped to scheduler time \(\tau\): syndrome-generation events occur at \(\tau=1,2,\dots\) (one cycle step each), while decoding-completion events are scheduled at \(\tau_{\text{finish}}=\tau_{\text{start}}+T_{\text{dec}}\).}

\subsubsection{Online Phase: Real-time Execution}

During execution, the LLI is streamed to quantum hardware, which generates a continuous stream of syndrome data corresponding to the executed operations. This stream is the primary input to Triage, which dispatches decoding tasks to a shared pool of \(M\) physical decoders. The scheduler is triggered by two events: (i) arrival of a new syndrome layer, which introduces new pending slices; and (ii) completion of a decoding task, which releases decoder capacity.

\revMulti{At each syndrome-arrival event, the engine checks whether synchronization constraints are satisfied for critical operations. If satisfied, execution proceeds normally. If not, the engine inserts an \emph{idle syndrome layer} at \(\ell\): all existing timeline layers with index \(t\ge \ell\) are shifted to \(t+1\), and a new layer \(t=\ell\) is created with per-qubit idle slices (while preserving spatial layout and carrying forward deadline semantics). This preserves causal order but delays all subsequent logical operations by one cycle. Decoders continue processing already assigned tasks asynchronously, and scheduling is retriggered after each decode completion and next syndrome arrival.} Therefore, the scheduler objective is to maximize decoder throughput and minimize the number of such idle-layer insertions.

\subsubsection{Problem Formulation}

Now we can formally define our task as a real-time scheduling problem. Given:
\begin{itemize}
    \item A shared pool of $M$ decoders, each with speed $r_{dec}^{i}$ relative to the syndrome generation speed.
    \item A dynamic, undirected constraint graph $G=(V, E)$ representing the set of all generated slices, slice attributes, and their mutually exclusive constraints.
\end{itemize}
The scheduler's task is to, at each decision point, find an assignment function $\pi: V' \to \{1, ..., M\}$ where $V'$ is a subset of all \texttt{PENDING} slices, satisfying:
\begin{enumerate}
    \item The chosen set of slices $V'$ must be an independent set in the graph $G$ (i.e., for any two slices $u, v \in V'$, there is no edge between them).
    \item The number of assigned slices cannot exceed the number of available decoders, $|V'| \le M_{available}$.
\end{enumerate}

The global objective is to produce a sequence of assignments that minimizes \emph{the total number of idle syndrome layers} when synchronizing the Pauli frames of critical operations, thus minimizing the overall logical error rate.

\subsection{The Dual-Mode Scheduling}
\label{sec:algorithm}

Triage combines a lightweight \emph{steady mode} for average-case throughput with a priority-aware \emph{emergency mode} that resolves causal cones for imminent critical operations.

\subsubsection{Steady Mode: Heuristic Scheduling}

At each syndrome-generation or decode-completion event, the scheduler selects up to $M_{available}$ conflict-free \texttt{PENDING} slices using a priority function $P(V)$. We explore several heuristic policies: First-In-First-Out (FIFO) prioritizes slices with the oldest timestamp to clear backlogs chronologically; Earliest-Deadline-First (EDF) prioritizes slices with the smallest deadline to proactively service operations closest to becoming critical; and Min-Degree-First (MDF) prioritizes slices with the fewest neighbors to minimize decoding latency.

To balance these critical factors, we propose a unified priority function:
\begin{equation}
    P(V) = w_u \cdot \text{Urgency}(V) + w_c \cdot \text{Cost-Efficiency}(V),
\end{equation}
where $w_u$ and $w_c$ are tunable weighting factors ($w_u + w_c = 1$). The urgency term quantifies proximity to a critical deadline, defined as $\text{Urgency}(V) = 1/\text{Deadline}(V)$. The cost-efficiency term favors slices that are computationally cheaper to decode, defined by the inverse of the slice's degree, $\text{Cost-Efficiency}(V) = 1/(\text{Degree}(V) + 1)$.

\begin{figure}[!ht]
    \centering
    \includegraphics[width=0.95\linewidth]{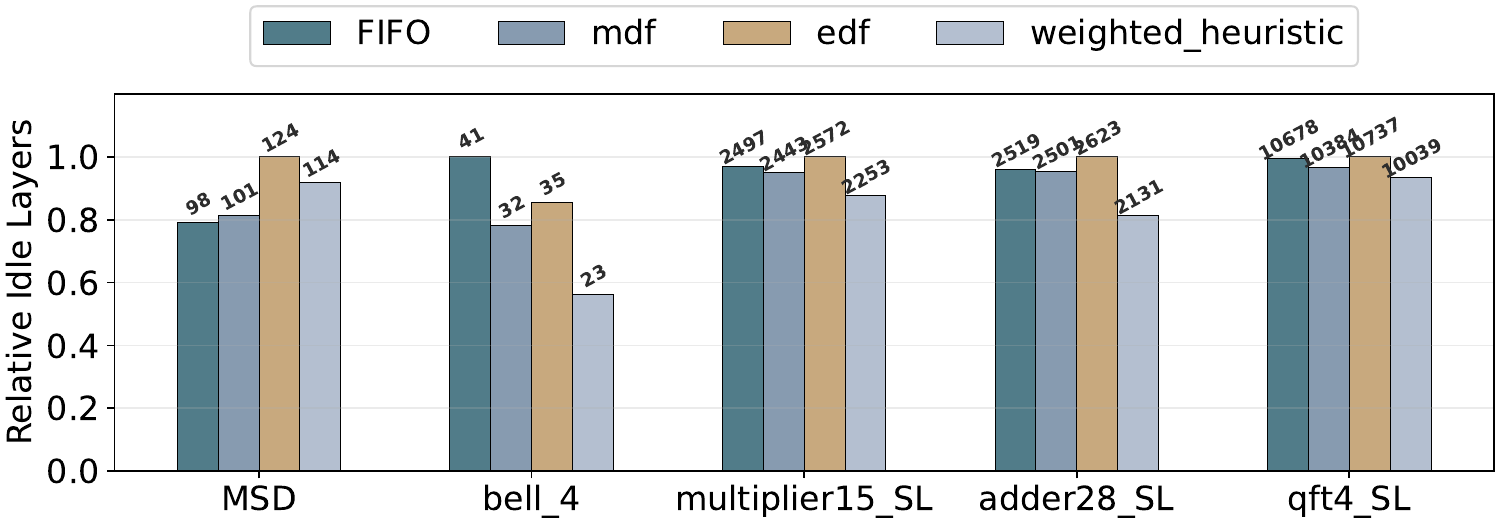}
    \caption{\revF{Relative idle layers inserted by different heuristic policies. Heuristic-only scheduling leaves significant room for improvement.}}
    \label{fig:heuristic_comparison}
\end{figure}

\revF{Figure~\ref{fig:heuristic_comparison} provides a preliminary evaluation of these policies (detailed setups are deferred to Section~\ref{sec:setup}). While our proposed weighted heuristic consistently outperforms the simple baselines, the sheer volume of idle layers remains substantial across all purely heuristic approaches. Pure heuristics inherently lack the foresight to guarantee low-latency Pauli frame updates for irregularly timed critical operations. This limitation motivates the emergency mode of our dual-mode architecture.}

\subsubsection{Emergency Mode: Predictive Causal Cone Coloring}

When the Triage Trigger signals an imminent deadline, the scheduler transitions to the emergency mode. Its objective is to resolve the causal cone of the impending critical operations with maximum parallelism, ensuring the necessary Pauli frames are updated before the Clifford correction executes.

Rather than making step-by-step decisions, the emergency mode employs a \emph{predictive coloring} algorithm, detailed in Algorithm~\ref{alg:emergency_plan}. This algorithm runs a discrete event simulation. It initializes a priority queue with only the \texttt{PENDING} slices in the on-demand causal cone, ensuring the input size minimal. The main loop advances a simulated clock to the next event and then greedily selects an independent set of tasks (Lines 10-14). The algorithm then records each selected slice in the final plan, and updates the auxiliary information (Lines 18-22). The core intuition is once inside the emergency mode, all slices in the causal cone share the same urgency. The primary factor for throughput is therefore the computational cost, resolved by the \emph{MDF} policy. The online scheduler then transitions to a simple executor, dispatching the pre-computed tasks from the plan at their scheduled start times.

\begin{algorithm}[!ht]
\caption{Predictive Causal Cone Coloring}
\label{alg:emergency_plan}
\begin{algorithmic}[1]
\State \textbf{Input:} Causal cone slice set $C$, current time $t_{now}$, decoder model $D_{model}$
\State \textbf{Output:} An emergency plan $P$
\State Initialize plan $P \gets \emptyset$
\State Initialize priority queue $Q$
\For{slice $s \in C$}
    \State $s.t_{start} \gets \max(t_{now}, s.t_{syndrome\_ready})$
    \State Push $s$ to $Q$, prioritized by $s.t_{start}$
\EndFor
\While{$Q$ is not empty}
    \State $t_{sim} \gets \text{NextEvent}(Q, D_{model})$ 
    \State $R \gets$ all slices from $Q$ where $s.t_{start} \le t_{sim}$
    \State Sort $R$ by degree
    \State $N_{free} \gets D_{model}$.num\_free($t_{sim}$)
    \State $D_{dispatch} \gets$ SelectConflictFree($R$, $N_{free}$)
    \For{slice $s \in D_{dispatch}$}
        \State Add $(t_{sim}, s)$ to $P$
        \State $t_{fin} \gets t_{sim} + \text{CalculateDuration}(s.\text{degree})$
        \For{neighbor $n \in Q$ of $s$}
            \State $n.t_{start} \gets \max(n.t_{start}, t_{finish})$
            \State $n.\text{degree} \gets n.\text{degree} - 1$
            \State Update position of $n$ in $Q$
        \EndFor
    \EndFor
    \State Re-insert non-dispatched slices from $R$ back into $Q$
\EndWhile
\State \textbf{return} $P$
\end{algorithmic}
\end{algorithm}

\textbf{\revB{Complexity Analysis}}
\revB{Let $n$ be the number of slices in the causal cone. The initialization (Lines 4-7) pushes $n$ elements into the priority queue $Q$, taking $O(n \log n)$ time. During the main loop, each slice is extracted and dispatched exactly once. Therefore, the inner loop (Lines 16-20) performs at most $6n$ neighbor updates, with each priority queue position update taking $O(\log n)$. Sorting the ready set $R$ (Line 10) is also bounded by $O(n \log n)$. Thus, the overall worst-case complexity scales efficiently at $O(n \log n)$. We will empirically validate this overhead in Section~\ref{sec:exp_overhead}.}

\subsubsection{The Triage Trigger}

The Triage Scheduler's adaptivity and efficiency is governed by the \emph{Triage trigger}, the mechanism that decides precisely when and how to transition to the emergency mode. The trigger is activated whenever any \texttt{PENDING} slice's deadline reaches a predefined threshold $\tau_{emergency}$ (e.g., $\tau_{emergency}=4$). \revMulti{To prevent the scheduling complexity from causing latency spikes on exceptionally large causal cones (which often accumulate near the end of highly entangled applications), we enforce a strict \texttt{ScopeCap}$<100$. If an evaluated causal cone exceeds this size limit, the scheduler falls back to the steady mode.}

To avoid thrashing, Triage re-plans only when all expansion-driven conditions hold:

\begin{itemize}
    \item The set of urgent slices introduces a causal cone that is not fully contained within the currently emergency scope.
    \item The expansion is significant, exceeding a defined fraction (e.g., 30\%) of the existing scope's size.
    \item A minimum time interval (e.g., 2) has passed since the last re-plan.
\end{itemize}

\begin{figure}[!ht]
    \centering
    \includegraphics[width=0.85\linewidth]{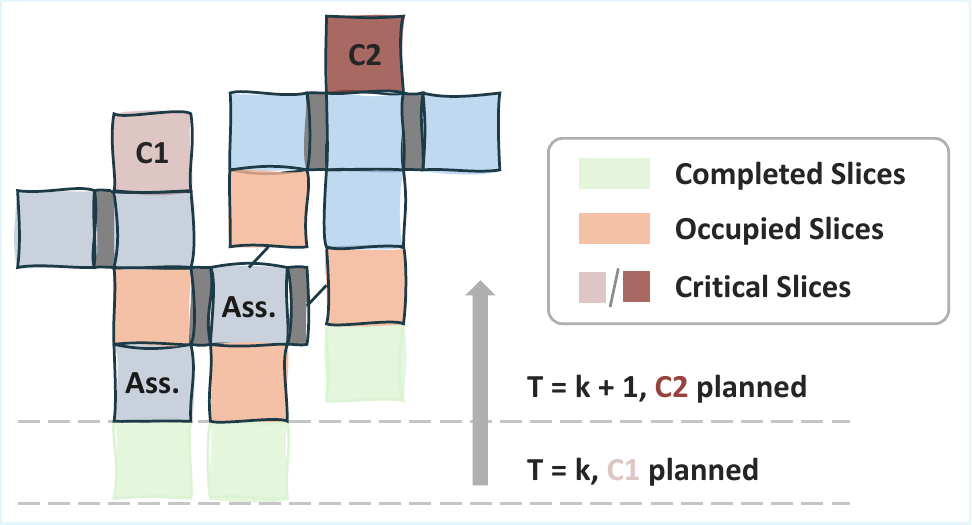}
    \caption{A 2-D snapshot of the Triage Trigger's operation. At $T=k+1$, a new critical slice $C_2$ triggers a scope expansion.}
    \label{fig:trigger_edge_cases}
\end{figure}

\revMulti{When expansion-driven conditions are met and there is overlap between the two scopes, the scheduler performs an incremental update.} Figure~\ref{fig:trigger_edge_cases} provides a 2-D simplified snapshot of this process. The scheduler operates on an evolving dependency graph. At time $T=k$, an emergency plan for a critical slice $C_1$ is already active. As time goes to $T=k+1$, a new critical slice $C_2$ becomes urgent, triggering a re-evaluation. The new emergency scope for $C_2$ excludes already \texttt{COMPLETED} slices, and the incremental planner will take into account the blocking effect of $C_1$'s plan on future slices.

\revShep{\textbf{Critical-path impact of scheduling.}
The part of Triage that can introduce noticeable latency overhead is emergency-mode causal-cone planning. Triage does not assume that every scheduling computation stalls the quantum processor. An emergency plan is cached and subsequently executed as a lightweight dispatch table. Therefore, only the portion of planning, dispatch, or interconnect latency that cannot be hidden behind ongoing decoding can affect the critical path. In Section~\ref{sec:exp_overhead}, we conservatively model this unhidden latency by delaying task start times.}

\subsubsection{Throughput Maximization via Opportunistic Backfilling}

\begin{figure}[!ht]
    \centering
    \includegraphics[width=\linewidth]{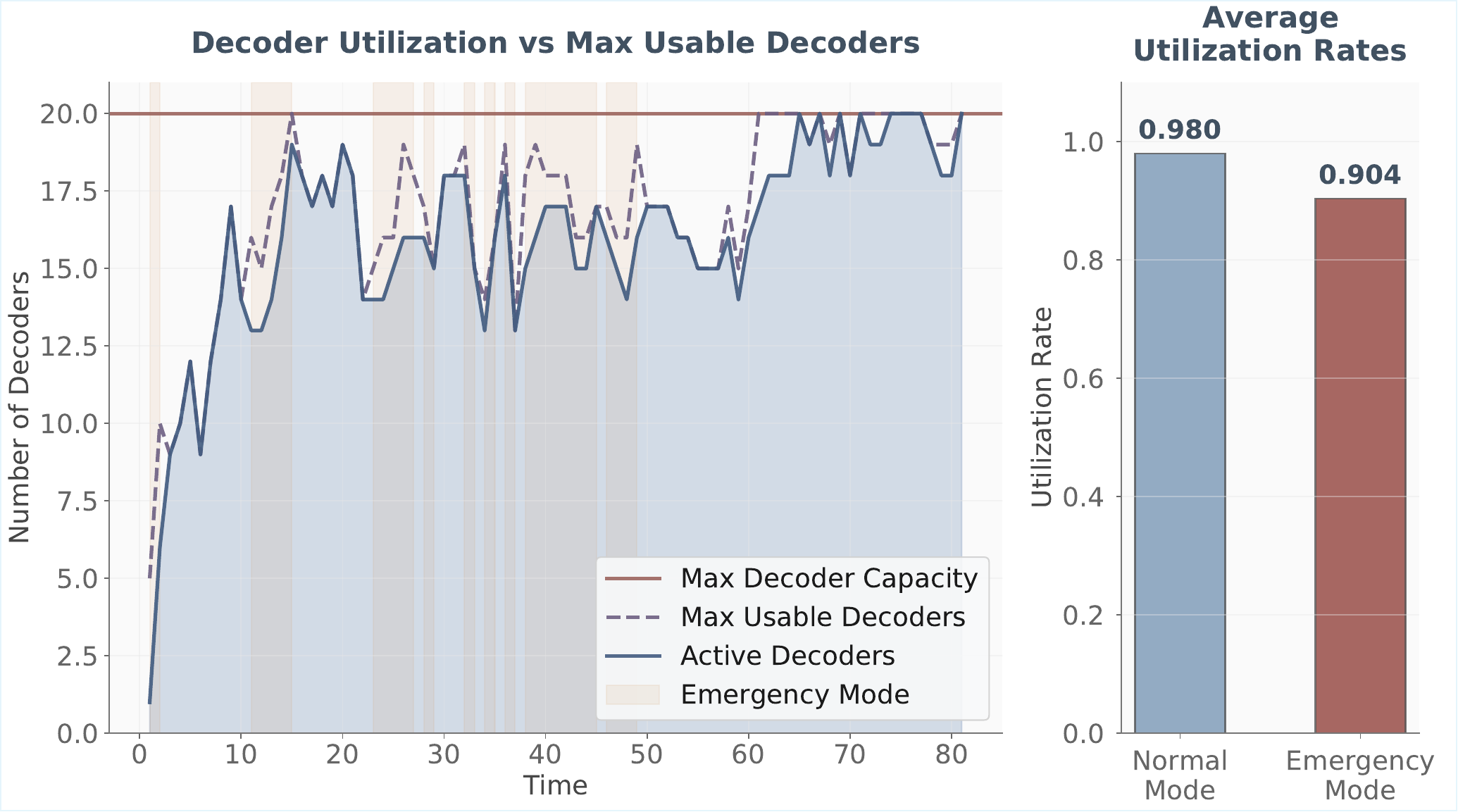}
    \caption{Motivation for Opportunistic Backfilling. Left: Triage's decoder utilization over time showing active decoders (blue), maximum capacity (purple dashed), and emergency periods (orange). Right: Comparison of utilization rates. The utilization can be improved by backfilling.}
    \label{fig:backfill_motivation}
\end{figure}

While the emergency mode is latency-optimal, its resource utilization can be inefficient. The parallelism of a causal cone often dictates a peak decoder requirement, $M_{peak}$, that is less than the total available decoders, $M$. As illustrated in Figure~\ref{fig:backfill_motivation}, this discrepancy creates idle decoders and wastes computational resources. To reclaim this lost throughput, we introduce an \emph{opportunistic backfilling} mechanism. \revF{The scheduler first computes $M_{peak}$ from the emergency plan, then derives the \emph{max usable decoders} for backfilling at each pass as
$M_{usable}(t)=\max\!\left(0,\min\!\left(M-M_{peak}-B_{bf}(t),\,F(t)-E(t)\right)\right)$, where $B_{bf}(t)$ is currently running backfill tasks, $F(t)$ is physically free decoders, and $E(t)$ is emergency tasks dispatched in the same pass.} This budget is then used to dispatch non-critical, causally-disconnected tasks using the heuristic scheduler, thereby maximizing throughput without any risk of interfering with the critical emergency plan.

\section{Experiment Setup}
\label{sec:setup}

\textbf{Simulation Framework.}
We develop a simulation framework that models the entire classical control pipeline: The compiler emits LLIs, the static analyzer constructs an annotated Timeline, and a discrete-event simulator generates syndromes and invokes the scheduler on syndrome arrivals and task completions. Before each critical operation, it checks whether the causal cone is decoded; otherwise it inserts an idle syndrome layer into the Timeline, and generates a layer of syndrome similar to the memory experiment~\cite{google2025quantum}. \revMulti{To prevent an unrecoverable backlog of decoding tasks, the simulation is forcibly terminated if the total number of inserted idle layers exceeds ten times the original layer count of the benchmark.} The scheduler is invoked on every syndrome generation and every task completion.

\textbf{Metrics.}
We evaluate scheduler performance using two metrics. Since an idle layer is inserted only when synchronization fails, we measure \emph{the number of inserted idle layers} as a direct metric for the scheduler's ability to handle critical operations. The simulation will also terminate when a significant backlog is detected.
The \emph{logical error rate (LER)} provides the ultimate measure which is correlated to the total execution layers.  We first simulate window-based lattice surgery using a circuit-level noise model, and then aggregate the LER of each layer to obtain the overall LER.

\textbf{System Configuration.}
We use a Litinski-style compiler~\cite{litinski2019game} to generate LLIs for our benchmarks. The instruction set is composed of multi-patch measurement, patch rotation and idle. To model the decoding time, we profiled the \texttt{pymatching} decoder~\cite{higgott2025sparse} on varying decoding volume, fitting the empirical data to a power-law model: $t_{decode} = A \cdot (\text{volume})^{\alpha}$. Given $\alpha=1.17$, our framework's decoding time for a given slice is determined by the size of its window buffer, which is directly related to the number of unresolved neighbors (i.e., its degree in the constraint graph). Note that our assumption that latency is monotonically increasing with volume holds for any practical decoder, so the relative performance trends in our evaluation are expected to be general. \revShep{Pattern-dependent runtime variation is modeled separately in Section~\ref{sec:stochastic_latency}, where a calibrated heavy-tail jitter model is injected into every decoder task.}

\revMulti{For Monte Carlo, we simulate a $d=9$ rotated surface code under circuit-level depolarizing noise at $p=3 \times 10^{-3}$, and we extrapolate the logical error rate to the $d=21$ case.} We perform the Monte Carlo simulations with Stim~\cite{gidney2021stim}, with each point made of at least $10^5$ runs.

\textbf{Benchmarks.} To cover a wide range of scenarios, we select a series of benchmarks from QASMBench~\cite{li2023qasmbench} with various T-gate densities. Furthermore, to demonstrate the universality of our framework, we include compiled versions for both Compact Layout (CL)~\cite{litinski2019game} and Standard Layout (SL)~\cite{hirano2025locality}. A summary of these benchmarks is provided in Table~\ref{tab:benchmarks}.

\begin{table*}[!t]
    \centering
    \caption{Characteristics of the FTQC Benchmark Suite.}
    \label{tab:benchmarks}
    \begin{tabular}{llcccc l}
        \toprule
        \revA{\textbf{Benchmark}} & \textbf{Short Name} & \textbf{\# LQubits} & \textbf{\# Layers} & \textbf{\# $T$-Gates} & \textbf{$T$-Den.$^{*}$} & \revA{\textbf{Category}} \\
        \midrule
        \revA{T-State Injection} & T\_injection & 9 & 13 & 1 & 7.69\% & \revA{FT Gadget} \\
        \revA{Arbitrary Rotation ($\pi/7$)} & rotation\_C+T & 1 & 2694 & 318 & 11.80\% & \revA{FT Benchmark} \\
        \revA{Magic State Distillation 15-to-1} & MSD15to1 & 5 & 24 & 11 & 45.83\% & \revA{FT Gadget} \\
        Bell State Preparation & bell4 & 4 & 41 & 5 & 12.20\% & \revA{FT Gadget} \\
        15-qubit Multiplier & mult15\_CL$^{*}$ & 15 & 586 & 252 & 43.00\% & \revA{Arithmetics} \\
        15-qubit Multiplier & mult15\_SL$^{*}$ & 15 & 508 & 252 & 49.61\% & \revA{Arithmetics} \\
        28-qubit Adder & adder28\_CL & 28 & 1894 & 168 & 8.87\% & \revA{Arithmetics} \\
        28-qubit Adder & adder28\_SL & 28 & 640 & 168 & 26.25\% & \revA{Arithmetics} \\
        \revA{64-bit Adder} & adder64\_SL & 64 & 1492 & 392 & 26.27\% & \revA{Arithmetics} \\
        \revA{118-bit Adder} & adder118\_SL & 118 & 2770 & 728 & 26.28\% & \revA{Arithmetics} \\
        11-qubit SECA & seca11\_SL & 11 & 140 & 56 & 40.00\% & \revA{Arithmetics} \\
        4-qubit Variational & variational4\_SL & 4 & 3636 & 402 & 11.06\% & \revA{Variational Algorithm} \\
        \revA{4-qubit QFT} & qft4\_SL & 4 & 1505 & 459 & 30.50\% & \revA{QFT Algorithm} \\
        \revA{4-qubit Trotterization} & trotter4\_SL & 4 & 2198 & 576 & 26.21\% & \revA{Hamiltonian Simulation} \\
        \revA{26-qubit Ising Model} & ising26\_SL & 26 & 11303 & 3688 & 32.63\% & \revA{Hamiltonian Simulation} \\
        \bottomrule
    \end{tabular}
\begin{flushleft}
$^{*}$ $T$-Den.: The proportion of $T$ gates. CL: Compiled with Compact Layout~\cite{litinski2019game}. SL: Compiled with Standard Layout~\cite{hirano2025locality}.
\end{flushleft}
\end{table*}

\textbf{Simulation Device. } \revMulti{All experiments were conducted with an Intel i9-14900K processor and 188 GB of RAM.} The simulation framework was implemented in Python 3.9.

\textbf{Baselines.}
While parallel window decoding has been extensively discussed~\cite{skoric2023parallel, tan2023scalable, viszlai2025swiper}, a framework for fine-grained scheduling has yet to be established. We construct baselines within our framework in Section~\ref{sec:model} to demonstrate the benefits of our spatio-temporal parallelism:
\begin{itemize}
    \item \emph{Serial sliding window~\cite{dennis2002topological}:} A scheduler processes a block of slices involved in a lattice surgery operation at a time, but does not process slices at later times in advance.
    \item \emph{Time-parallel window~\cite{skoric2023parallel}:} A scheduler leverages parallelism across the time dimension for logical patches, but does not split up multi-qubit operations.
    \revMulti{\item \emph{SWIPER~\cite{viszlai2025swiper}:} A state-of-the-art speculative scheduler. We reproduce its \emph{successor-based} strategy which is optimistic regarding mis-speculations, setting 10\% misprediction rate and 10\% speculation time. Furthermore, the speculative decoding module is not included in the decoder usage.}
\end{itemize}

\section{Evaluation Results}

We design our evaluation based on the following key questions to reflect the practicality of our framework.

\begin{itemize}
    \item[\textbf{Q1}] How does the spatio-temporal parallelism compare to default and SOTA strategies under varying constraints?
    \item[\textbf{Q2}] How do the proposed schedulers perform across a diverse suite of FTQC applications in terms of idle reduction and logical error rates?
    \item[\textbf{Q3}] \revMulti{How resilient is the Triage scheduler to real-world decoder latency fluctuations?}
    \item[\textbf{Q4}] \revMulti{What are the computational overheads of the proposed schedulers? Does Triage's advantage degenerate when considering scheduling and interconnect latency?}
    \item[\textbf{Q5}] How do internal mechanisms and hyperparameters contribute to the overall performance?
\end{itemize}

In the following simulations, the heuristic weights are set to $w_u=w_c=0.5$, the Triage trigger's replan scope threshold is 0.3, and the minimum planning interval is 2.

\subsection{Motivating Spatio-Temporal Parallelism}

\begin{figure}[!t]
    \centering
    \subfloat[Fixed decoding speed of 0.8.\label{fig:idle_vs_decoders}]{\includegraphics[width=0.47\linewidth]{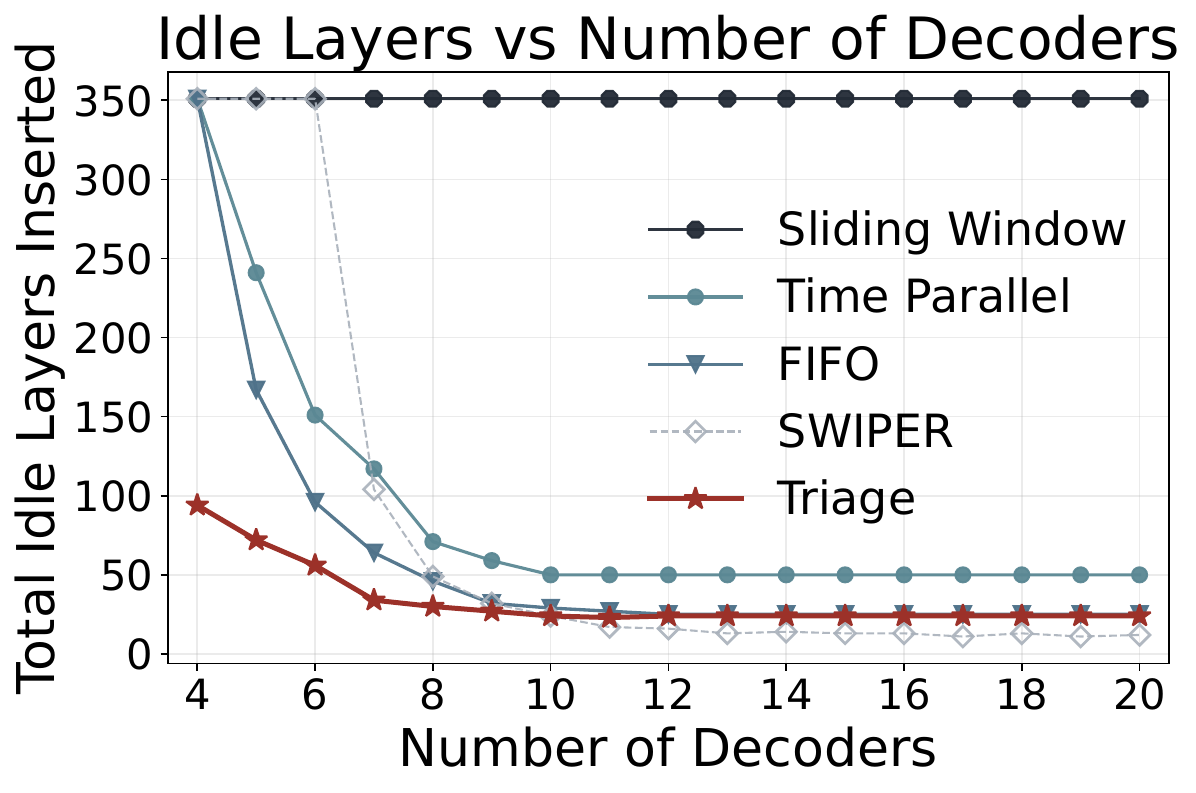}}
    \hfill
    \subfloat[Fixed pool of 8 decoders.\label{fig:idle_vs_speed}]{\includegraphics[width=0.47\linewidth]{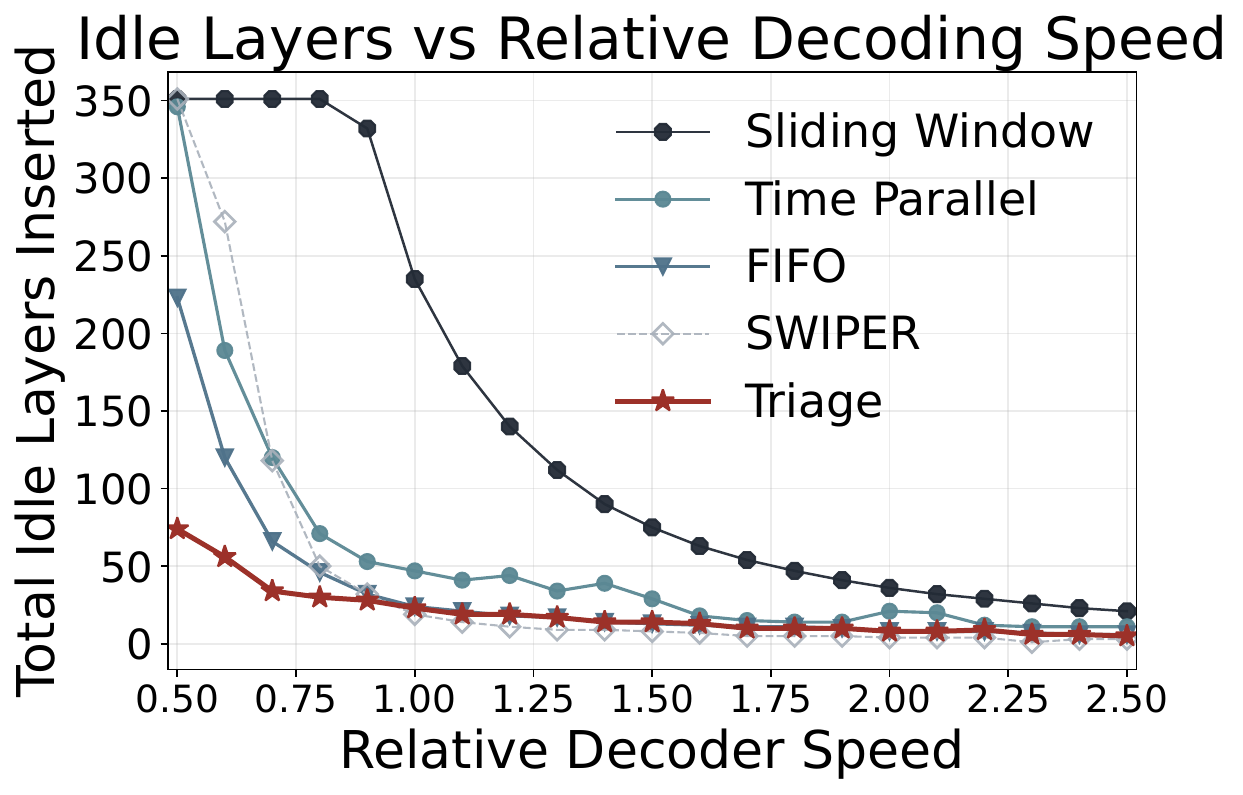}}
    \caption{\revMulti{Relation between idle layers inserted and (a) number of available decoders; (b) relative decoding speed ($\tau_{dec} / \tau_{gen}$).}}
    \label{fig:motivation_charts}
\end{figure}

We first evaluate five schedulers representing a spectrum of parallelization strategies on the Bell4 on Litinski's compact layout. This task comprises 39 logical layers and includes 5 critical $\pi/8$ gates. The schedulers under comparison are: the baseline \emph{sliding window} and \emph{time-parallel} schedulers; \revMulti{the speculative scheduler \emph{SWIPER}~\cite{viszlai2025swiper};} our \emph{time-space-parallel} scheduler with FIFO policy; and our \emph{Triage} scheduler. Figure~\ref{fig:motivation_charts} shows the number of inserted idle layers as we vary the number of available decoders and the relative speed of each individual decoder.

\paragraph{Observation 1: Serial processing is fundamentally unscalable}
As shown in both figures, the \emph{sliding window} scheduler exhibits the worst performance. The flat line demonstrates that its sequential nature makes it fundamentally unable to leverage parallel hardware resources to increase throughput.

\paragraph{Observation 2: Spatio-temporal parallelism enables superior resource utilization}
The \emph{time-parallel} scheduler offers a significant improvement over the serial approach, but its performance saturates at a high number of idle layers. The \emph{time-parallel} scheduler is bottlenecked by its inability to break down correlated multi-qubit operations, resulting in a high floor. In contrast, the \emph{time-space-parallel} schedulers can process these complex operations with a much finer granularity, achieving a lower saturation point.

\paragraph{\revMulti{Observation 3: Triage outperforms SWIPER under resource constraints}}
\revMulti{\emph{SWIPER} leverages its speculative mechanism to achieve extremely high parallelism, showing competitive performance when resources are abundant. However, these advantages diminish significantly in resource-constrained regimes where speculative overheads can lead to resource contention. In contrast, our \emph{Triage} scheduler demonstrates superior performance in these scenarios.}

\subsection{Design Space Exploration of Scheduling Strategies}

\begin{figure*}
    \centering
    \includegraphics[width=1\linewidth]{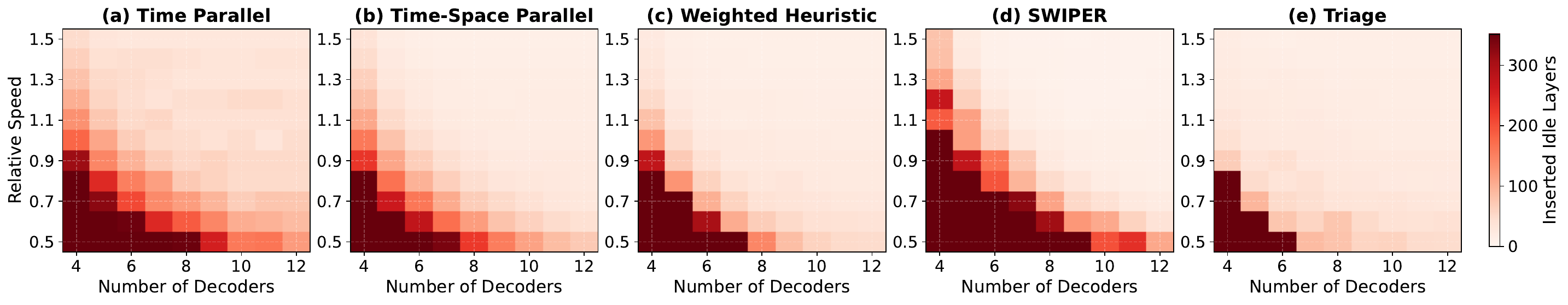}
    \caption{\revMulti{Heatmaps illustrating the number of inserted idle layers for different schedulers across various decoder counts and relative speeds. Darker red indicates a higher number of idle layers, signifying worse performance. The \emph{Triage} scheduler consistently achieves near-best performance across the entire space and defines the performance frontier in resource-constrained scenarios.}}
    \label{fig:heatmaps_all}
\end{figure*}

\revMulti{To characterize the performance landscape of various scheduling strategies, we conduct a design space exploration across a wide range of decoder counts and relative speeds, comparing our \emph{Triage} scheduler against the \emph{Time Parallel} scheduler, the \emph{Time-Space Parallel} (FIFO) scheduler, and the \emph{SWIPER}~\cite{viszlai2025swiper}}. Figure~\ref{fig:heatmaps_all} presents the performance of these schedulers as heatmaps, where darker regions indicate a higher number of inserted idle layers and thus poorer performance. The \emph{Triage} scheduler's performance is particularly pronounced in the most challenging regions where decoders are both slow (low y-axis value) and scarce (low x-axis value), whereas \revMulti{\emph{SWIPER} achieves the global minimum of idle layers when decoding resources are relatively abundant.}

\begin{figure}[!ht]
    \centering
    \includegraphics[width=\linewidth]{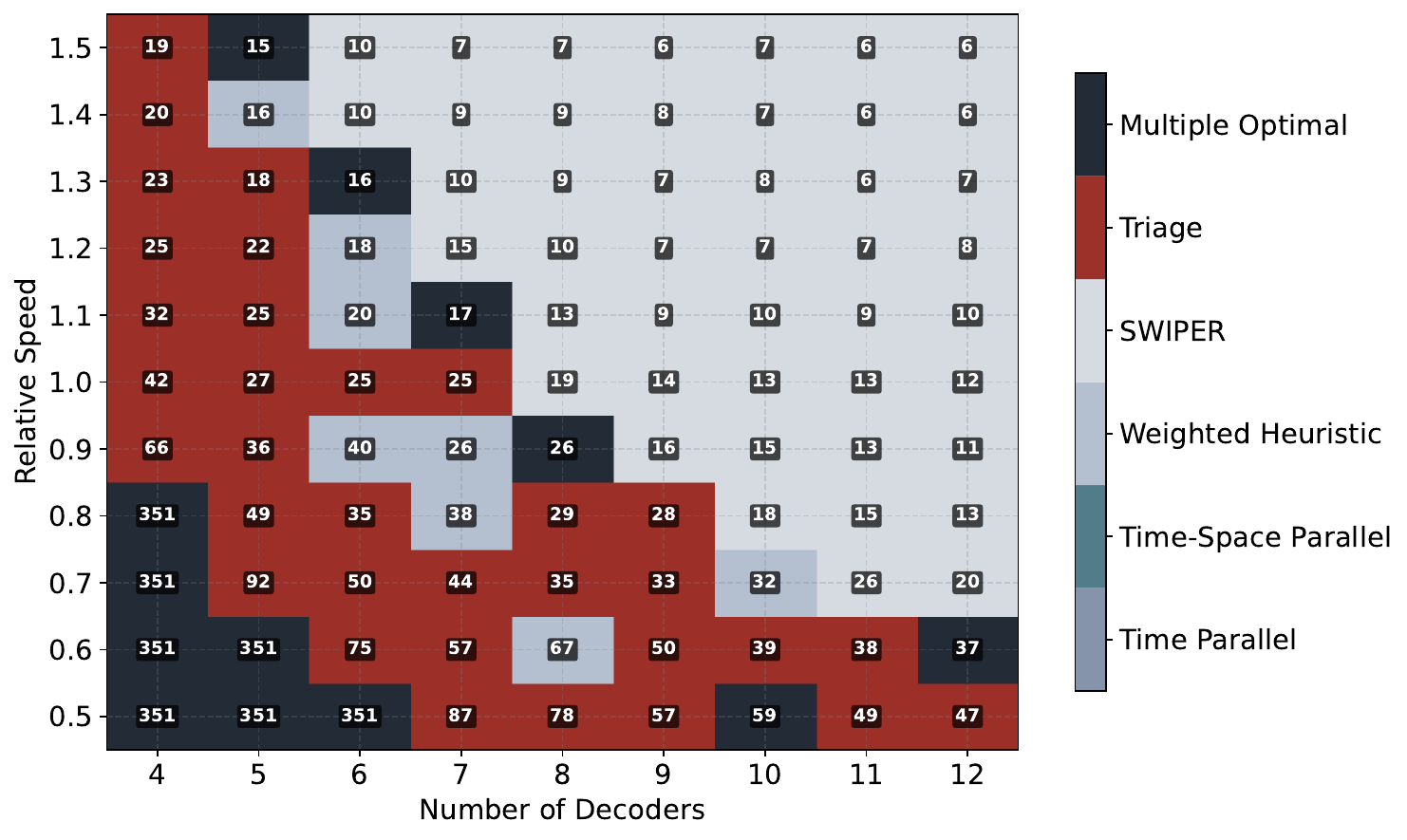}
    \caption{\revMulti{The optimal scheduler map on the Bell4 application. Each cell in the grid is colored according to the best-performing scheduler for that decoder pool configuration.}}
    \label{fig:optimal_scheduler_map}
\end{figure}

Figure~\ref{fig:optimal_scheduler_map} synthesizes these results into an optimal map. Each cell is colored to indicate which scheduler achieved the best performance for that specific resource configuration. 
\revShep{The \emph{Triage} scheduler (red) defines most of the feasible resource-constrained lower-left frontier.} In contrast, \emph{SWIPER} (light blue) tends to be optimal in the resource-abundant upper-right regime. \revE{  Notably, the lower-left black regions denote a regime of failure where extreme resource scarcity forces all schedulers to trigger the backlog-induced termination threshold.}

\subsection{Performance Across Benchmarks}

\begin{figure*}[!ht]
    \centering
    \subfloat[Slow-Decoder Scenario (2$\times$ decoders, speed=0.9)\label{fig:bench_slow}]{\includegraphics[width=0.85\textwidth]{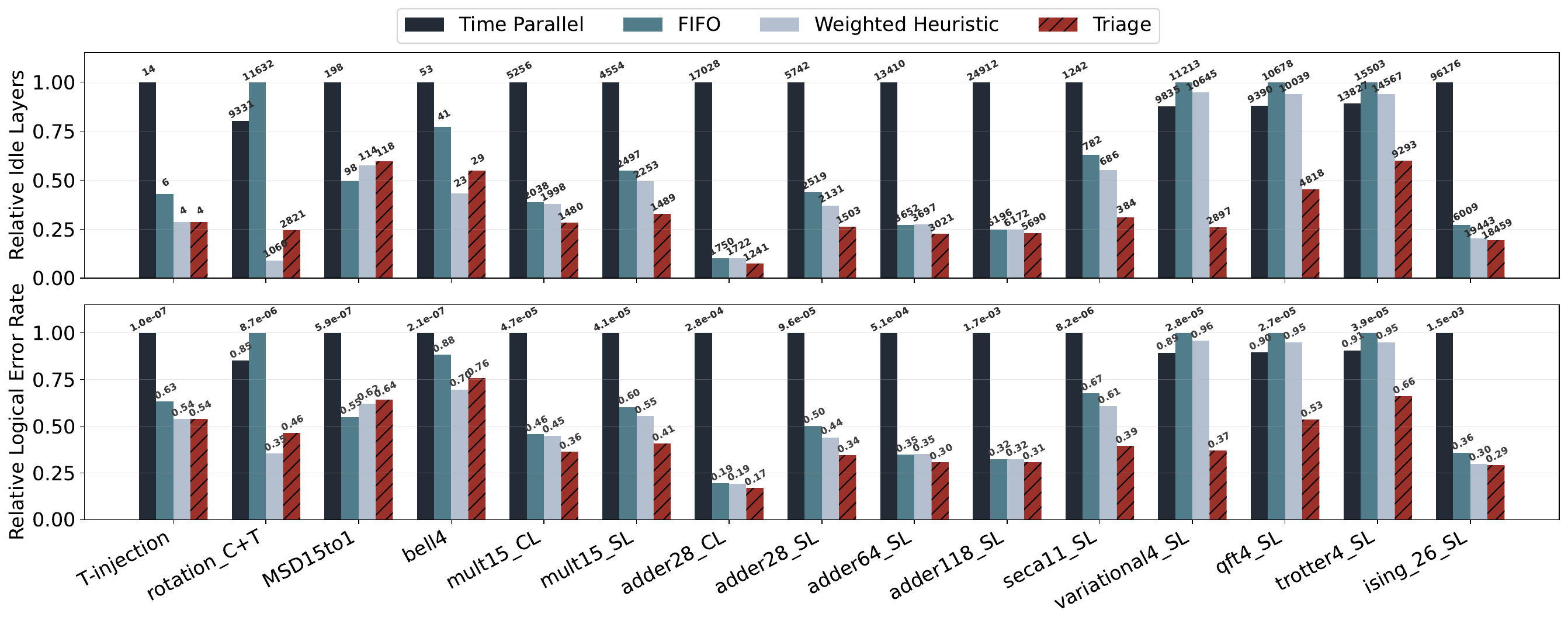}}
    \\[0.5ex]
    \subfloat[Fast-Decoder Scenario (1$\times$ decoders, speed=1.8)\label{fig:bench_fast}]{\includegraphics[width=0.85\textwidth]{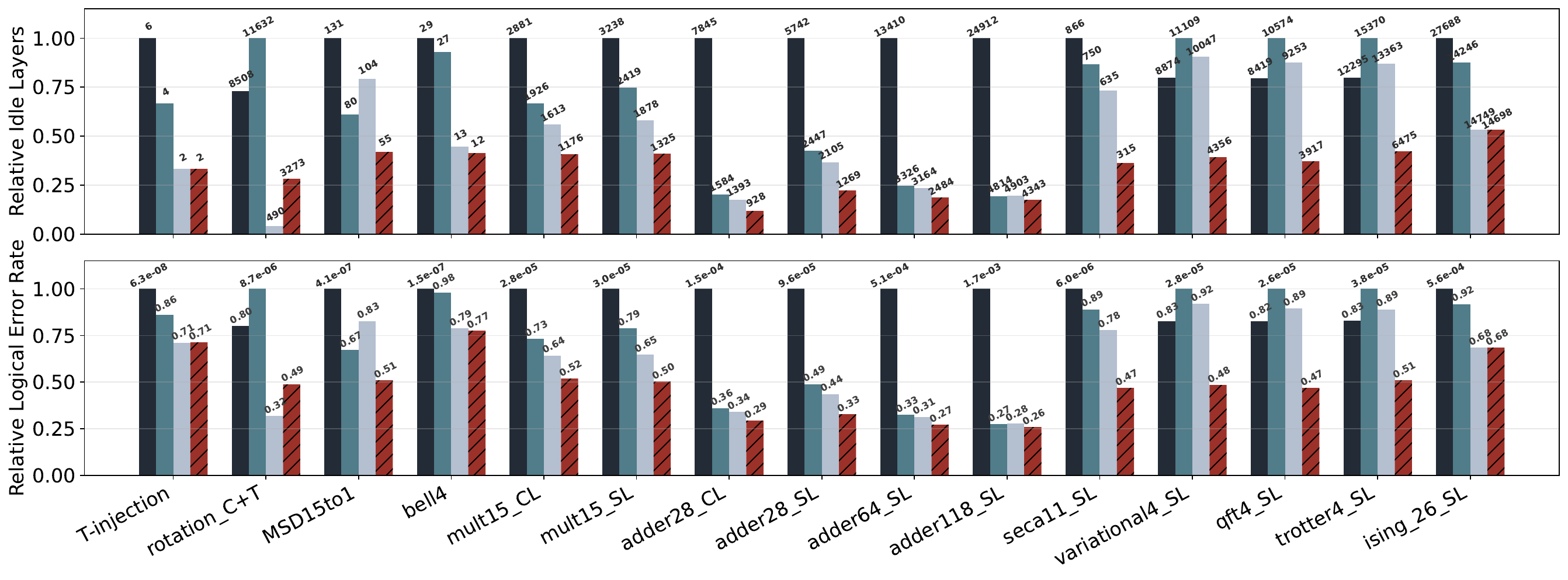}}
    \caption{\revMulti{LER comparison across all benchmarks for (a) a resource-constrained scenario and (b) a resource-abundant scenario. Lower bars indicate better performance. Triage outperforms the baseline in nearly all cases.}}
    \label{fig:benchmark_charts}
\end{figure*}

We now evaluate the schedulers on various FTQC benchmarks. Figure~\ref{fig:benchmark_charts} illustrates the \revMulti{idle layers inserted} and LER across all benchmarks in two representative resource scenarios: a \textbf{Parallelism-Rich Scenario}, featuring numerous but slow decoders (count = 2$\times$\#LQs, speed=0.9), and a \textbf{Latency-Rich Scenario}, featuring few but fast decoders (count = \#LQs, speed=1.8). \revMulti{The height of the idle-step bars is normalized within each application, while the absolute values are labeled.}

The results in both Figure~\ref{fig:bench_slow} and Figure~\ref{fig:bench_fast} show a trend of hierarchy of performance. \revMulti{For visual clarity, \emph{SWIPER} is omitted as its performance under resource-constrained scenarios is comparable to the FIFO policy.} The \emph{time-only parallelism} scheduler performs the worst, while the \emph{Triage} scheduler consistently achieves the best or near-best performance. The FIFO policy itself is not particularly bad, as starting from the bottom of the timeline results in most allocated slices having small degrees. Across these benchmarks, Triage achieves an average logical error rate reduction of 52.6\% compared to the time-parallel baseline.

An intriguing exception is the variational algorithm, where time-space parallel scheduling using a single strategy mode performs worse than time-only parallel scheduling alone. In this case, splitting the multi-qubit logical operations into dependent slices increases scheduling difficulty, yet the performance of the Triage scheduler remains superior. 

\revE{\textbf{Estimation of Physical Execution Time.}
The total wall-clock time is determined by the total number of layers (including inserted idles) and the duration of each layer: $T_{total} = N_{total\_layers} \times T_{layer}$. Each logical layer in a surface code typically requires $d$ rounds of syndrome measurements, so $T_{layer} = d \times T_{meas}$. For a distance $d=21$ code, $T_{layer}$ varies significantly across platforms: approximately 21~$\mu$s for superconducting qubits ($T_{meas} \approx 1\mu$s), and ranging from 2.1~ms to 21~ms for ion traps or neutral atoms. By reducing the number of idle layers our \emph{Triage} scheduler translates directly into significant wall-clock time savings.}

\subsection{\revB{Impact of Stochastic Decoding Latency}}
\label{sec:stochastic_latency}

\revB{In practical FTQC systems, decoding latency is not deterministic but fluctuates due to varying error patterns. We model decoder latency jitter as a mean-preserving lognormal factor,
\begin{equation}
t_{\text{actual}} = t_{\text{estimated}}\cdot \exp\!\left(-\frac{\sigma^2}{2}+\sigma z\right),\quad z\sim\mathcal N(0,1),
\end{equation}
so that the mean latency remains consistent while introducing a heavy tail characteristic of real-time systems.
The jitter scale \(\sigma\) is parameterized as
\begin{equation}
\begin{split}
    \sigma(d, p) = \text{clamp}( & \sigma_{base} + \alpha_d \log_2(d/5) \\ 
    & + \alpha_p(p - p_{ref}), \sigma_{min}, \sigma_{max} )
\end{split}
\end{equation}
Here, \(\sigma_{\text{base}}\) is the baseline jitter at \((d=5,p=p_{\text{ref}})\), \(\alpha_d\) captures distance-driven complexity growth, and \(\alpha_p\) captures error-rate-driven complexity growth.}

\revShep{The calibration set is built from per-shot \texttt{pymatching} latency measurements on Stim-generated rotated surface-code circuits with 15K measured shots per setting after warmup. We obtain $\sigma_{\text{base}}=0.3447$, $\alpha_d=0.0041$, $\alpha_p=15.03$, $p_{\text{ref}}=10^{-3}$, $\sigma_{\min}=0.30$, $\sigma_{\max}=0.70$. Leave-one-out validation predicts the held-out \(\sigma\) with a mean absolute error of 0.064 and captures tail quantiles with about 15\% relative error. Thus, the lognormal model is a calibrated heavy-tail service-time abstraction used to test whether Triage remains robust when complex syndrome patterns create tail latency.}

\begin{figure*}[!ht]
    \centering
    \begin{minipage}[c]{0.60\textwidth}
        \centering
        \subfloat[Noiseless vs. Noisy LER across benchmarks.\label{fig:noisy_main}]{\includegraphics[width=\linewidth]{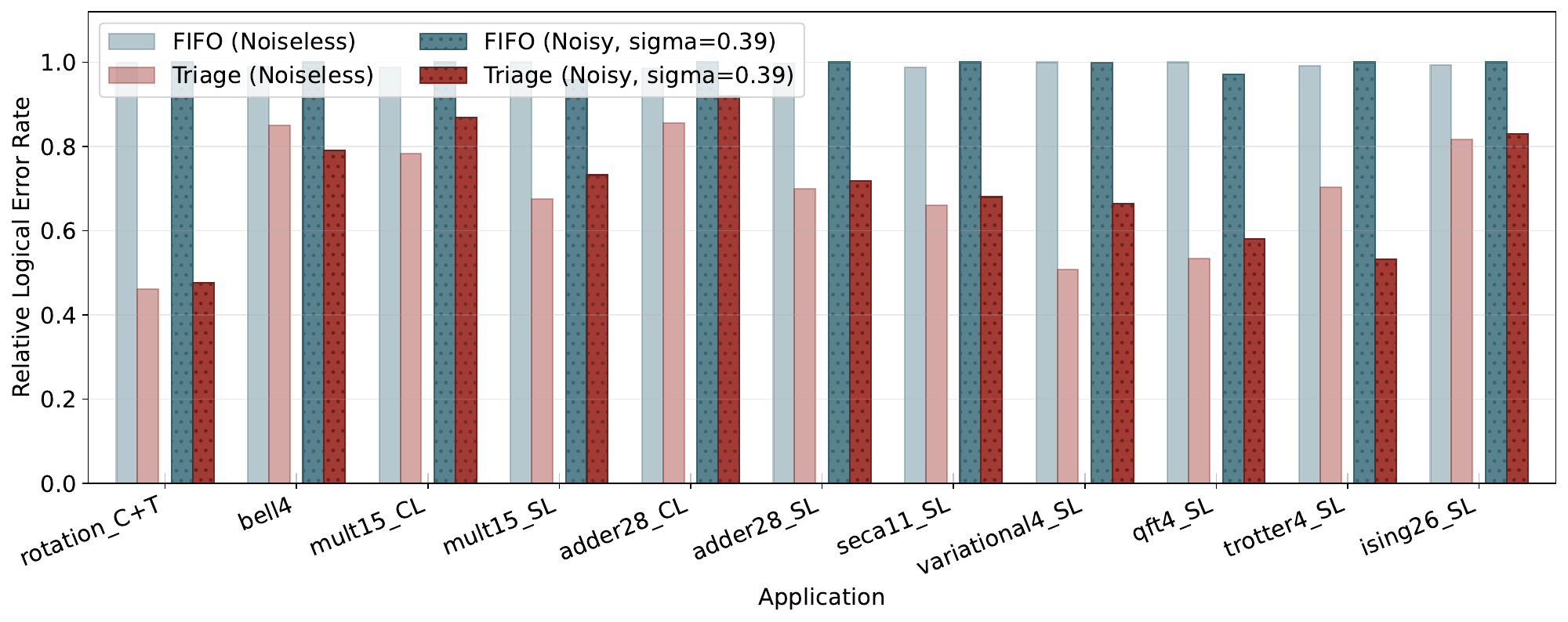}}
    \end{minipage}
    \hfill
    \begin{minipage}[c]{0.36\textwidth}
        \centering
        \subfloat[Empirical Lognormal $\sigma$: Measured vs. Predicted.\label{fig:sigma_fit}]{\includegraphics[width=\linewidth]{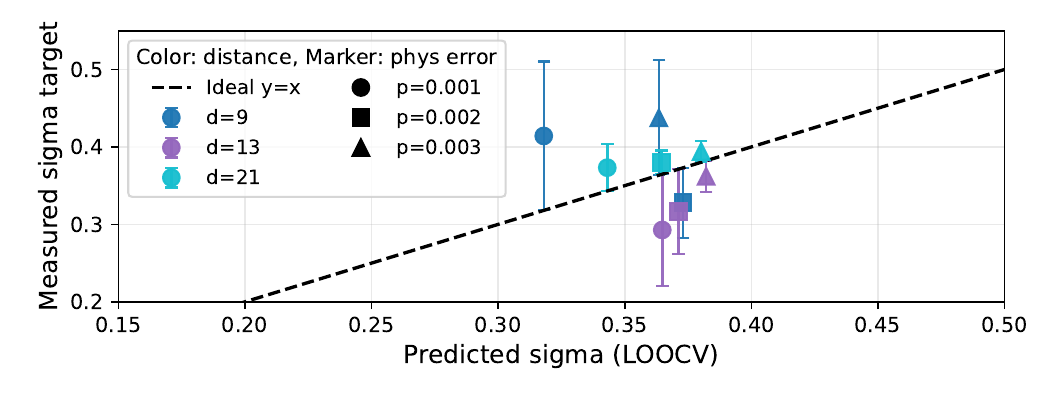}}
        \\[0.6ex]
        \subfloat[LER sensitivity to $\sigma$ (Multiplier15\_SL).\label{fig:sigma_sensitivity}]{\includegraphics[width=\linewidth]{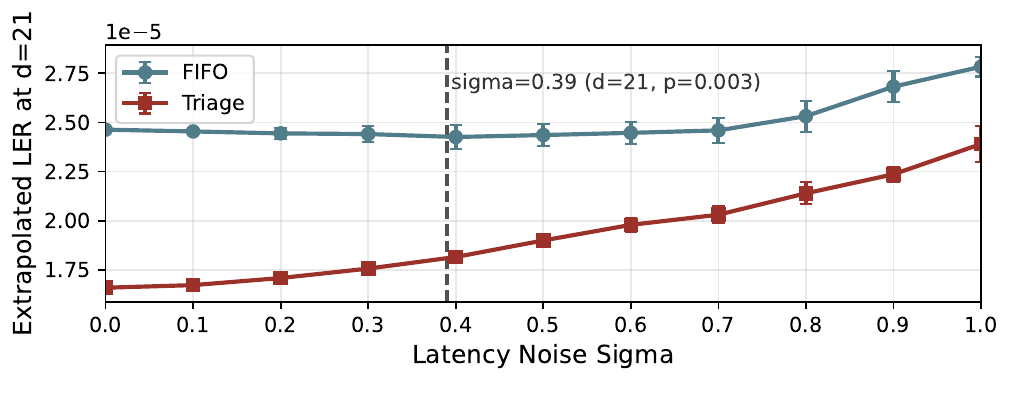}}
    \end{minipage}
    \caption{\revB{Evaluation of the Triage scheduler under stochastic latency. (a) compares LER in noiseless and noisy scenarios; (b) validates our $\sigma(d,p)$ fit against empirical measurements; (c) demonstrates Triage's robustness as latency jitter increases.}}
    \label{fig:stochastic_eval}
\end{figure*}

\revB{As shown in Figure~\ref{fig:sigma_fit}, our lognormal model closely matches the empirical $\sigma$ measured from \texttt{pymatching}. Figure~\ref{fig:noisy_main} presents a detailed comparison of LER for each application under both noiseless and noisy environments. Although the presence of stochastic latency inevitably leads to a higher LER, \emph{Triage} consistently maintains a significant advantage over the baseline. This robustness is further quantified in Figure~\ref{fig:sigma_sensitivity}, which tracks the LER of the Multiplier15\_SL benchmark as the jitter intensity $\sigma$ varies from 0 to 1. While the gap between FIFO and Triage narrows at extreme noise levels, \emph{Triage} remains the superior strategy.}

\subsection{\revMulti{Computational Overhead Analysis}}
\label{sec:exp_overhead}

\begin{figure*}[!ht]
    \centering
    \subfloat[Total scheduling time and latency]{\includegraphics[width=0.51\linewidth]{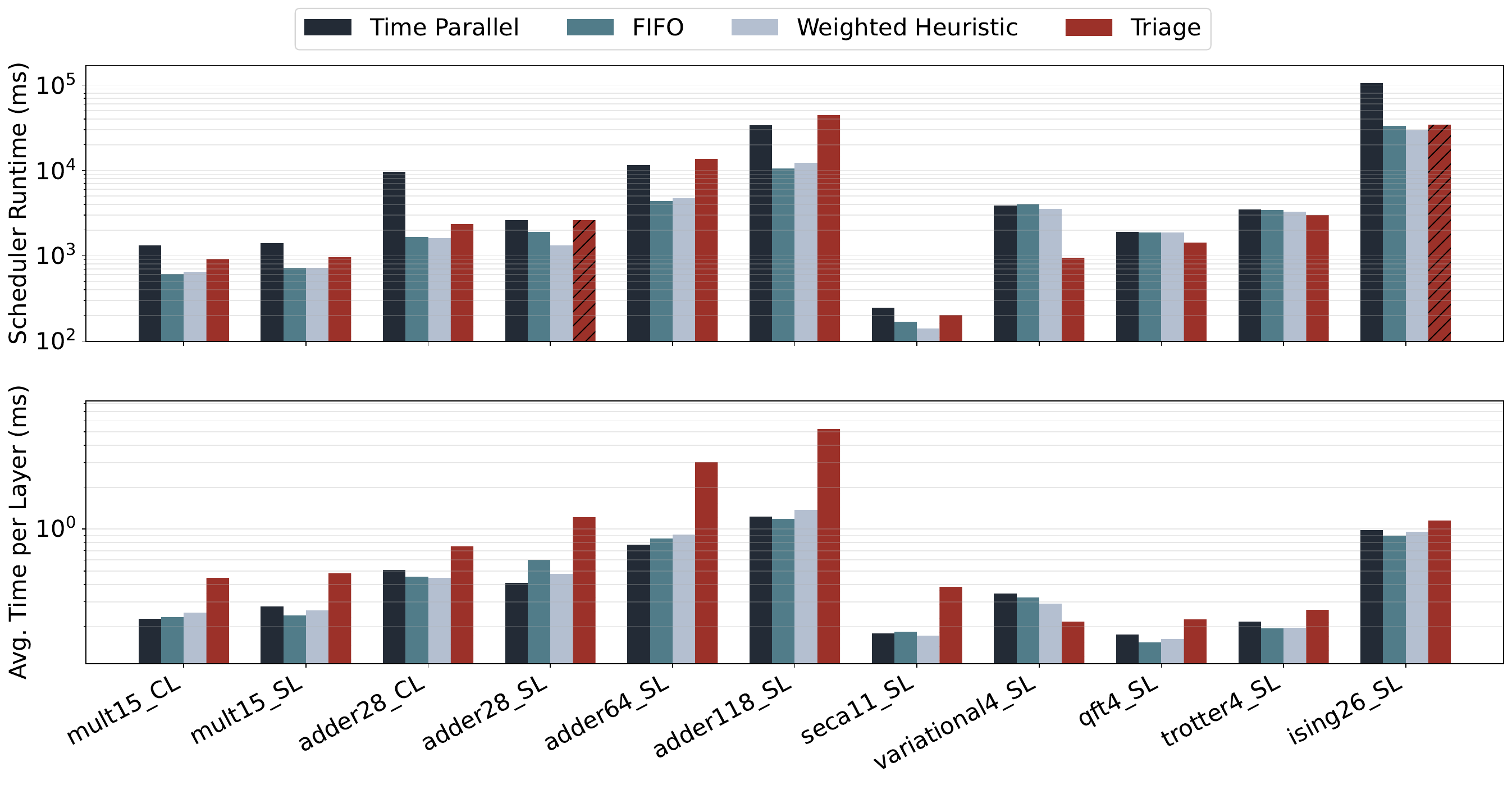}\label{fig:runtime_a}}
    \hfill
    \subfloat[Plan time vs. scope size]{\includegraphics[width=0.19\linewidth]{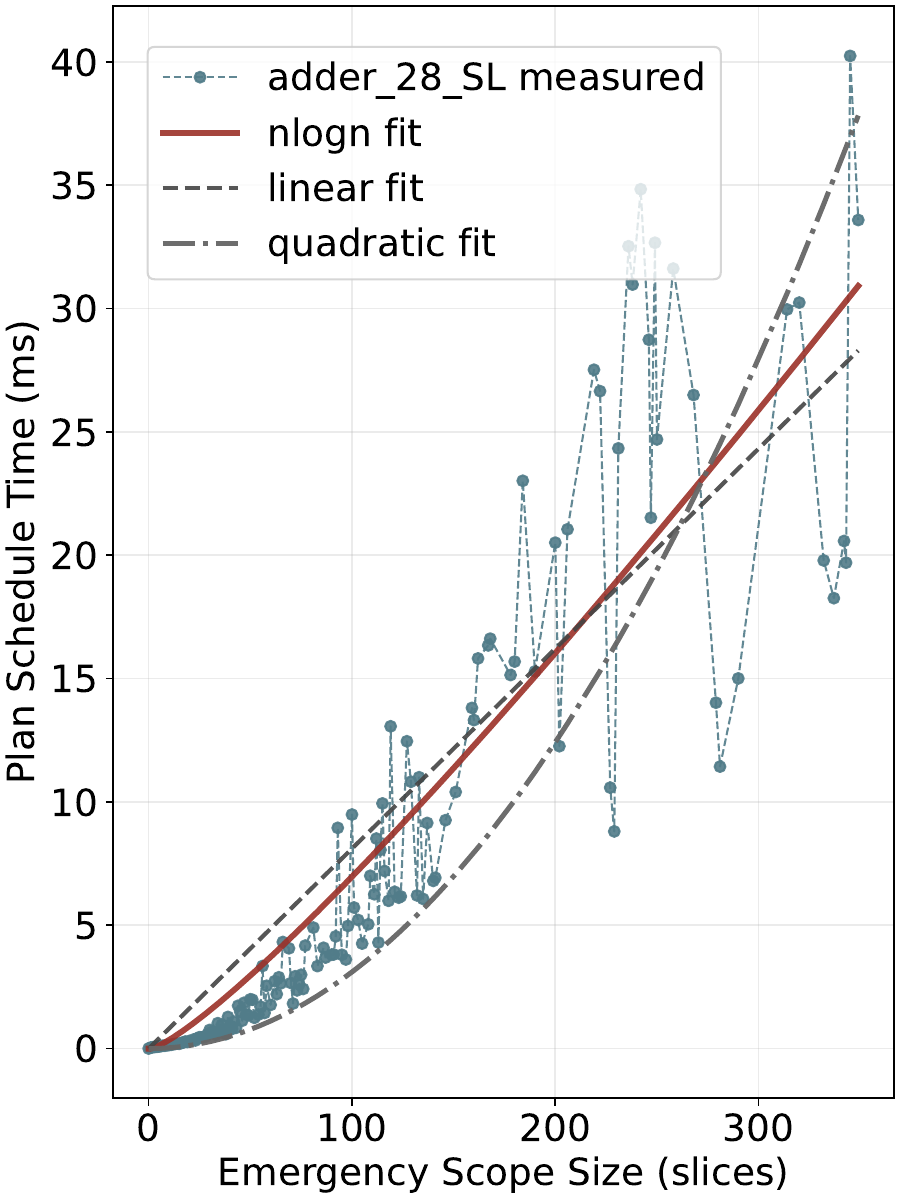}\label{fig:runtime_b}}
    \hfill
    \subfloat[Performance under latency]{\includegraphics[width=0.25\linewidth]{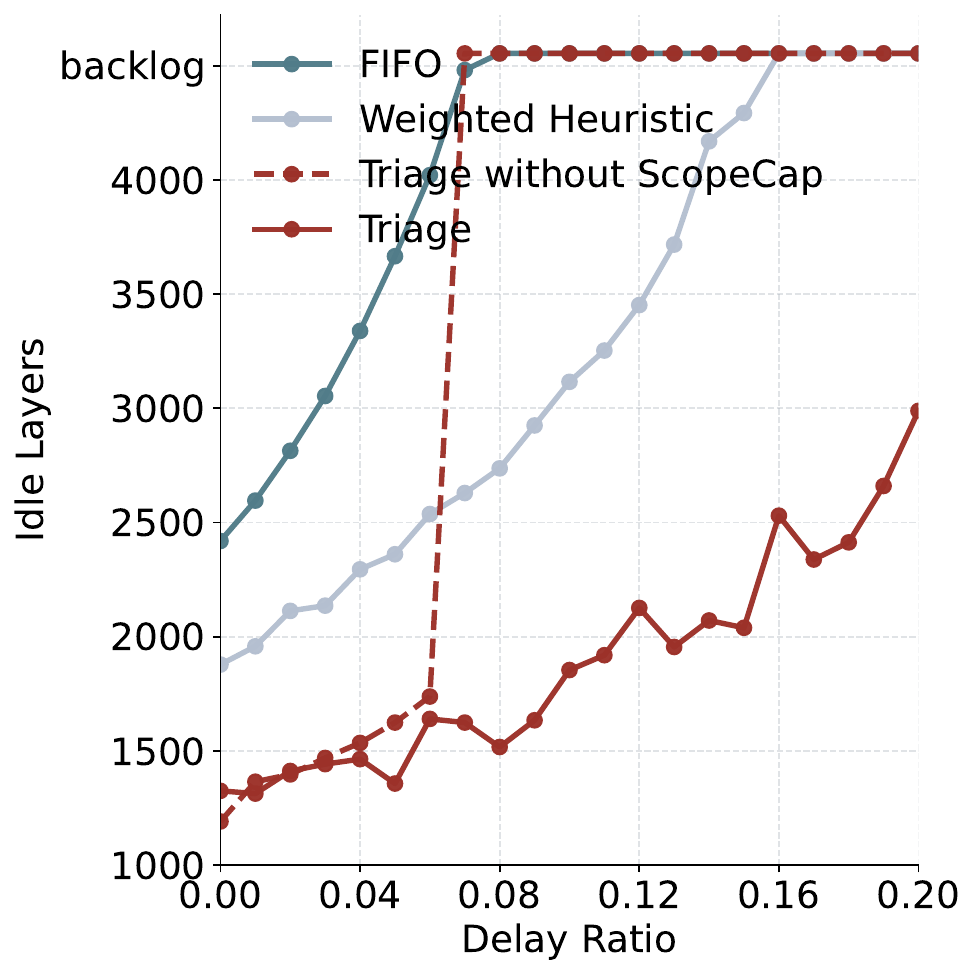}\label{fig:runtime_c}}
    \caption{\revMulti{Computational overhead analysis. (a) Total scheduling time per application (top) and average latency per logical layer (bottom). (b) Plan schedule time versus emergency scope size, best captured by an $O(n\log n)$ fit ($y=a\cdot n\log n$, $a=0.01513$, $R^2=0.805637$). (c) System performance (Idle Layers) under simulated scheduler latency.}}
    \label{fig:runtime_analysis}
\end{figure*}

Figure~\ref{fig:runtime_a} breaks down the runtime into total scheduling time and average latency per logical layer. Note that all reported runtimes strictly isolate the pure algorithmic latency required to generate a scheduling plan, excluding the simulation environment. \revShep{The measured wall-clock numbers characterize our Python prototype. Triage has sub-millisecond median per-layer scheduling cost across our benchmarks, but large emergency scopes can create multi-millisecond tail latency. This is acceptable for slow-cycle platforms such as ion traps or neutral atoms, but a superconducting implementation with $\sim$20~$\mu$s logical-layer cycles would require a compiled or hardware-assisted implementation. Our claim is that the algorithmic structure is bounded: emergency planning scales as $O(n\log n)$ and ScopeCap prevents pathological causal cones from entering the critical path.}

\revMulti{To rigorously assess how this scheduler latency impacts overall system performance, we conduct a sensitivity simulation that incorporates real-time scheduling delays. We define a \emph{Delay Ratio} as the baseline FIFO scheduler runtime relative to the decoding time, sweeping this ratio from 0.00 to 0.20. For heuristic policies, task assignment is delayed proportionally. For \emph{Triage's} emergency mode, the delay is dynamically calculated using the fitted $O(n\log n)$ function based on the real-time scope size. As shown in Figure~\ref{fig:runtime_c} with the Multiplier15\_SL benchmark, the \emph{Triage without ScopeCap} policy suffers from severe runtime penalties when attempting to resolve massive causal cones near the end of applications, causing the system to hit a backlog failure at a delay ratio of 0.06. By enforcing the \emph{ScopeCap}, \emph{Triage} maintains robust and superior performance across the entire delay spectrum.}

\subsection{Sensitivity and Ablation Studies}

\subsubsection{Impact of Decoding Window Size}
The decoding window size presents a critical trade-off between decoding throughput and individual operation fidelity. Smaller windows accelerate processing but reduce syndrome context, whereas larger windows improve accuracy at the cost of increased latency. This balance is dictated by classical resource availability.

\begin{figure}[!ht]
    \centering
    \includegraphics[width=0.75\linewidth]{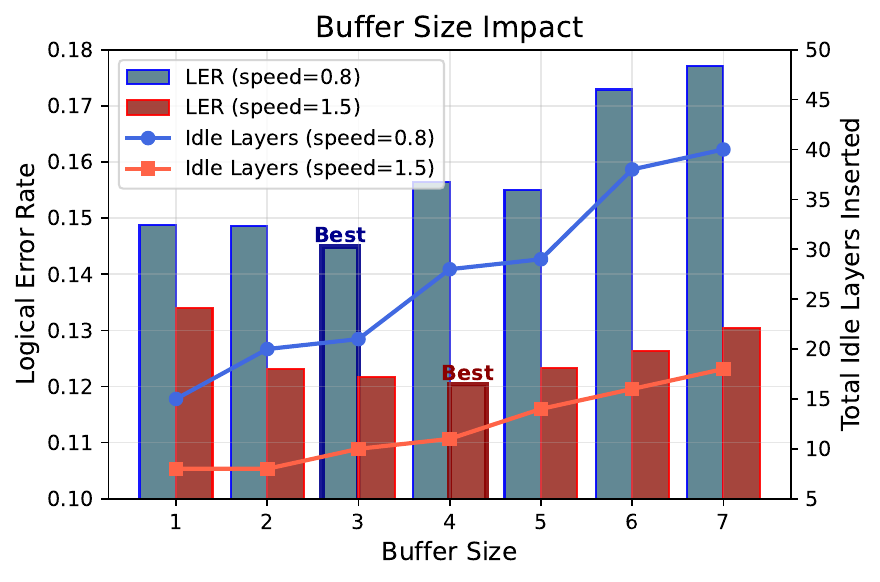}
    \caption{LER (bars) and inserted idle layers (lines) as a function of the window decoding buffer size with Triage at $d=7$. The decoder count is fixed to 8. An optimal operating point appears around a buffer size of $d/2$.}
    \label{fig:buffer_tradeoff}
\end{figure}

Figure~\ref{fig:buffer_tradeoff} illustrates this trade-off under two regimes: resource-constrained (speed$=0.8\times$) and resource-rich (speed$=1.5\times$). In the constrained regime, smaller buffers are optimal as they maintain high decoding throughput, thereby minimizing total idle layers and the resulting aggregate LER. Conversely, in resource-rich scenarios, the bottleneck shifts from throughput to individual operation fidelity, favoring larger windows. Consequently, the optimal buffer size is hardware-dependent; while current latency-limited systems necessitate smaller buffers, future high-performance hardware will likely favor larger windows approaching $d/2$~\cite{lin2025spatially}.

\subsubsection{Impact of Hyperparameters}

\begin{figure}[!ht]
    \centering
    \subfloat[Heuristic weight]{\includegraphics[width=0.48\linewidth]{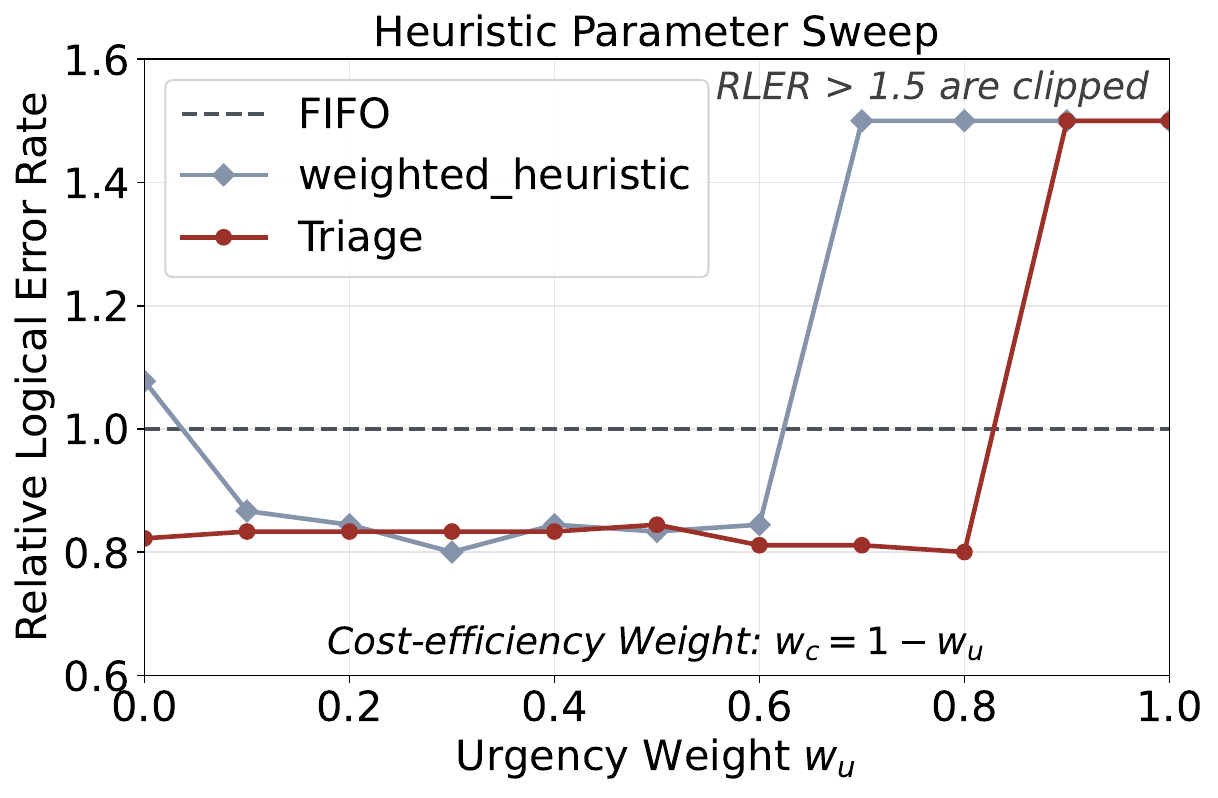}\label{fig:hyper_weight}}
    \hfill
    \subfloat[Emergency threshold]{\includegraphics[width=0.48\linewidth]{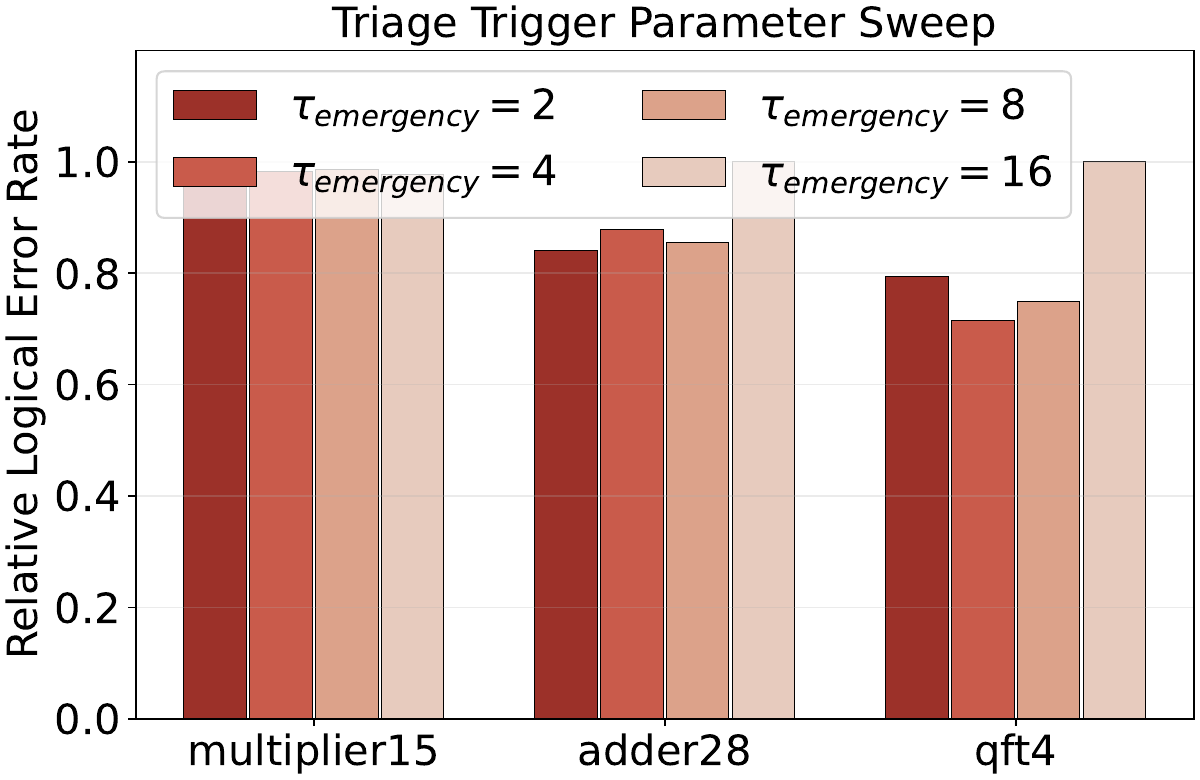}
    \label{fig:hyper_trigger}}
    \caption{Sensitivity analysis of \emph{Triage}. Triage has robust performance across a wide range of parameter configurations.}
    \label{fig:hyperparameters}
\end{figure}

\revMulti{We evaluate \emph{Triage}'s sensitivity to its primary hyperparameters: the heuristic weight ($w_u$) and the emergency threshold ($\tau_{emergency}$).} As Figure~\ref{fig:hyperparameters}(a) shows, sweeping $w_u$ from 0 to 1 reveals that the logical error rate is remarkably robust, eliminating the need for application-specific tuning. \revMulti{Similarly, Figure~\ref{fig:hyperparameters}(b) demonstrates stable performance across a moderate threshold range ($\tau_{emergency} \in [2, 8]$). Performance degrades only when excessively high thresholds (e.g., $\tau_{emergency} = 16$) delay necessary interventions during congestion spikes.} Overall, \emph{Triage} exhibits strong robustness to parameter variations without requiring fine-grained tuning.

\section{Related Work}

\textbf{Improvements on Decoders.} Research on decoders for FTQC has focused on improving accuracy, latency, and scalability. This includes algorithmic approaches, such as lookup table decoders~\cite{tomita2014low, das2022lilliput, ryan2021realization}, minimum-weight perfect matching (MWPM) decoders for surface codes~\cite{dennis2002topological, fowler2013minimum, higgott2022pymatching}. In addition, system-level approaches have been investigated to reduce decoding latency~\cite{kim2024fault, alavisamani2024promatch, vittal2023eraser, liao2023wit, thantharate2023q}. These include specialized solutions for superconducting qubits~\cite{holmes2020nisq, ueno2021qecool, higgott2025sparse, vittal2023astrea}, hierarchical decoders~\cite{delfosse2020hierarchical}, optimized union-find decoders~\cite{das2022afs}, and FPGA-based implementations~\cite{liyanage2023scalable, liyanage2024fpga}. These individual-decoder efforts complement our framework, which schedules a shared decoder pool to manage system-wide latency constraints.

\textbf{Decoder Scheduling.} Most existing works on decoder design assume a dedicated decoder is statically allocated for each logical qubit~\cite{bombin2023modular, skoric2023parallel, tan2023scalable}. \revMulti{Recent concurrent works have begun to address the challenges of dynamic decoder scheduling~\cite{bombin2023modular, maurya2024managing, viszlai2025swiper}, with \emph{SWIPER}~\cite{viszlai2025swiper} representing the current SOTA. Our framework focuses on mitigating the decoding pressure induced by non-Clifford operations, achieving lower logical error rates under resource-constrained scenarios.} Furthermore, our simulation enables a direct evaluation of how classical resource bottlenecks dictate final performance.

\textbf{Compilers for Optimizing Lattice Surgery.} Many compilers have been proposed to improve the scheduling of lattice surgery operations~\cite{lao2018mapping, molavi2025dependency, watkins2024high, leblond2023realistic, herr2017optimization, zhu2026o3ls}, and several works have also focused on increasing the parallelism during these procedures~\cite{beverland2022assessing, beverland2022surface, hamada2024efficient, hirano2025locality}. In our study, we use the compiler introduced in~\cite{watkins2024high, leblond2023realistic} to compare different strategies for decoder scheduling. Integrating advanced compiler techniques may further improve overall performance.

\section{Conclusion}

In this work, we identified the management of classical decoder resources as a bottleneck for scalable FTQC. \revA{We utilized a spatio-temporal framework to formulate the constrained dynamic scheduling problem.} We then proposed \emph{Triage}, a dual-mode scheduling architecture that maximizes resource utilization. Our implementation focused on surface codes, the principles of constrained parallel-window scheduling are broadly relevant. Extending this framework to general QLDPC codes will be a promising future work. Furthermore, exploring the co-design of the quantum compiler and scheduler represents a next step, enabling the compiler to optimize circuits with classical resource awareness.


\section*{Acknowledgment}
We would like to thank the anonymous reviewers for their helpful feedback and suggestions.

\IEEEtriggeratref{57}
\bibliographystyle{IEEEtran}
\bibliography{ref}

\end{document}